\documentclass[fleqn,usenatbib]{mnras}

\usepackage{newtxtext,newtxmath}

\usepackage[T1]{fontenc}

\DeclareRobustCommand{\VAN}[3]{#2}
\let\VANthebibliography\thebibliography
\def\thebibliography{\DeclareRobustCommand{\VAN}[3]{##3}\VANthebibliography}

\usepackage{graphicx}	
\usepackage{amsmath}	
\usepackage{multirow}
\usepackage{subcaption}
\usepackage{placeins}
\usepackage{soul}

\title[Spectral analysis of GRGs]{A spatially-resolved spectral analysis of giant radio galaxies with MeerKAT}

\author[K.K.L. Charlton et al.]{K.K.L. Charlton,$^{1}$\thanks{E-mail: chrkat009@myuct.ac.za}
J. Delhaize,$^{1}$
K. Thorat,$^{2}$
I. Heywood,$^{3,4,5}$
M. J. Jarvis,$^{3,6}$
M. J. Hardcastle, $^{7}$
Fangxia An,$^{8,9}$  \newauthor 
I. Delvecchio,$^{10}$
C. L. Hale,$^{3}$ 
I. H. Whittam,$^{3,6}$ 
M. Br\"{u}ggen,$^{11}$
L. Marchetti,$^{1,12}$
L. Morabito,$^{13}$ \newauthor
Z. Randriamanakoto,$^{14,15}$ 
S. V. White$^{13,5}$ 
A.R. Taylor,$^{1,6,16}$
\\
$^{1}$Department of Astronomy, University of Cape Town, Private Bag X3, Rondebosch 7701, South Africa\\
$^{2}$Department of Physics, University of Pretoria, Private
Bag X20, Hatfield 0028, South Africa\\
$^{3}$ Oxford Astrophysics, Denys Wilkinson Building, University of Oxford, Keble
Road, Oxford OX1 3RH, UK \\
$^{4}$Department of Physics and Electronics, Rhodes University, PO Box 94, Makhanda 6140, South Africa\\
$^{5}$ South African Radio Astronomy Observatory, 2 Fir Street, Black River Park, Observatory, Cape Town 7925, South Africa \\
$^{6}$ Department of Physics and Astronomy, University of the Western Cape, Robert Sobukwe Road, Bellville 7535, South Africa\\
$^{7}$Department of Physics, Astronomy and Mathematics, University of Hertfordshire, College Lane, Hatfield AL10 9AB, UK\\
$^{8}$Purple Mountain Observatory, Chinese Academy of Sciences, 10 Yuanhua Road, Qixia District, Nanjing 210023, People\'s Republic of China \\
$^{9}$Inter-University Institute for Data Intensive Astronomy, Department of Physics and Astronomy, University of the Western Cape \\
$^{10}$ INAF - Osservatorio di Astrofisica e Scienza dello Spazio di Bologna, via Gobetti 93/3, I-40129 Bologna, Italy\\
$^{11}$Hamburger Sternwarte, University of Hamburg, Gojenbergsweg 112, D-21029 Hamburg, Germany \\
$^{12}$INAF - Istituto di Radioastronomia, via Gobetti 101, I-40129 Bologna, Italy \\
$^{13}$Centre for Extragalactic Astronomy, Department of Physics, Durham University, Durham DH1 3LE, UK \\
$^{14}$South African Astronomical Observatory, PO Box 9, Observatory 7935, Cape Town, South Africa \\
$^{15}$Department of Physics, Faculty of Sciences,University of Antananarivo, B.P. 906, Antananarivo 101, Madagascar \\
$^{16}$The Inter-University Institute for Data Intensive Astronomy (IDIA), Department of Astronomy, University of Cape Town \\
}

\date{Accepted XXX. Received YYY; in original form ZZZ}

\pubyear{2023}

\begin{document}
\label{firstpage}
\pagerange{\pageref{firstpage}--\pageref{lastpage}}
\maketitle

\begin{abstract}
In this study we report spatially resolved, wideband spectral properties of three giant radio galaxies (GRGs) in the COSMOS field: MGTC J095959.63+024608.6 , MGTC
J100016.84+015133.0 and MGTC J100022.85+031520.4. One such galaxy MGTC J100022.85+031520.4 is reported here for the first time with a projected linear size of 1.29 Mpc at a redshift of 0.1034. Unlike the other two, it is associated with a brightest cluster galaxy (BCG), making it one of the few GRGs known to inhabit cluster environments. We examine the spectral age distributions of the three GRGs using new MeerKAT UHF-band (544-1088 MHz) observations, and $L$-band (900-1670 MHz) data from the MeerKAT International GHz Tiered Extragalactic Exploration (MIGHTEE) survey. We test two different models of spectral ageing, the Jaffe-Perola and Tribble models,  using the Broadband Radio Astronomy Tools (\textsc{brats}) software which we find agree well with each other. We estimate the Tribble spectral age  for MGTC J095959.63+024608.6 as 68 Myr, MGTC J100016.84+015133.0 as 47 Myr and MGTC J100022.85+031520.4 as 67 Myr.  We find significant disagreements between these spectral age estimates and the estimates of the dynamical ages of these GRGs, modelled in cluster and group environments. Our results highlight the need for additional processes which are not accounted for in either the dynamic age or spectral age estimations.
\end{abstract}

\begin{keywords}
radio continuum: galaxies -galaxies: active –galaxies: evolution –galaxies: nuclei.
– methods: observational
\end{keywords}



\section{Introduction}

\label{ch: Intro}

Giant radio galaxies (GRGs) display jets and lobes of sychrotron-emitting plasma that extend greater than 700 kpc in projected linear length, representing the most extended population of radio galaxies (\citealt{1974Natur.250..625W, schoenmakers2000new, 2023JApA...44...13D}). Their extensive relativistic jets and lobes may play an important role in the evolution of their host galaxy and surrounding environment via `jet-mode' active galactic nuclei (AGN) feedback \citep{2006MNRAS.370..645B, 2006MNRAS.365...11C, 2012ARA&A..50..455F, 2014ARA&A..52..589H}. In this mode of feedback, the AGN plasma jets travel through the inter-stellar medium (ISM), disturbing and heating the surrounding gas. This is thought to prevent gas cooling and stop accretion onto the super massive black hole (SMBH), halting its growth. In GRGs, since the jets are so large, this process extends further into the intergalactic and intercluster medium (IGM and ICM). Thus, GRGs could be ideal probes of the impact of AGN activity on the IGM and may provide insight into the nature of the environment itself. \citep{2008MNRAS.385.2117S,safouris2009mrc, 2013MNRAS.432..200M,2015MNRAS.449..955M}  \\

The reason for the really large sizes of GRGs is still uncertain. An early suggestion was made by \cite{1998A&A...329..431M} that the extended size of GRGs is due to the under-densities found within the IGM where the speed of the plasma is not as eroded by interactions with intervening matter. However, recent studies have reported that between 4-10\% of identified GRGs reside within high-density cluster environments \citep{Debhade...2020b,2020MNRAS.499...68T, 2024arXiv240308037S}. Another potential explanation for the occurrence of GRGs is that these galaxies contain abnormally powerful AGN engines which allow the jets to expand to large scales in short amounts of time \citep{article}. Counter to this, \citet{2009ARep...53.1086K} and \citet{2019A&A...622A..12H} found no evidence of the massive linear size of GRGs being due to particularly powerful central engines.\\

A more likely scenario then is that GRGs represent the oldest AGN systems, where the radio jets have had enough time to evolve and grow \citep{subrahmanyan1996morphologies}.  This scenario was previously thought to be implausible because the extrapolated number densities of GRGs as of 2021 were too low; only $\sim$820 had been identified compared to the hundreds of thousands of reported normal radio sources \citep{Debhade...2020b}. However, recent work (e.g., \citealt{2021MNRAS.501.3833D,2024arXiv240308037S, oei2023measuring, 2024arXiv240500232M}) suggests that the actual GRG sky density is significantly higher than indicated by previous studies. These works suggest that the surface brightness sensitivity of telescopes was actually the limiting factor, since these giants are often faint and diffuse. According to \citet{2024arXiv240308037S}, who performed a large-scale study of the radio, optical, and infrared properties of GRGs compared to regular radio galaxies, age is the main factor determining a radio galaxy's length. Given enough time for the galaxy to evolve, the other main factors in the late-stage size of a radio galaxy are the jet power, the host galaxy mass, and the large-scale environment. \\

If GRGs do represent the oldest AGN systems, then studying their extended structure is crucial to understanding the evolution of galaxies, their interactions, and their effect on the IGM. This is now becoming feasible, with the number of identified GRGs increasing dramatically in the last couple of years. The most notable of these is due to the second data release of the LOFAR Two-metre Sky Survey (LOTSS-DR2, \citealt{LOFARDR2}) and the work of \citet{2024arXiv240500232M}, increasing the total number of identified GRGs in the literature to $\sim$ 12000. \\

Complementary to LOFAR, which operates at $\sim$144 MHz, the MeerKAT radio interferometer in South Africa \citep{2016mks..confE...1J} is an excellent instrument for the study of GRGs at higher frequencies ($\sim$ 1 GHz). This is due to its excellent point source sensitivity, similar resolution capabilities, and excellent $uv$ coverage. It can currently observe in three different bands: the UHF (544-1088 MHz) and $L$ (856-1670 MHz) bands, as well as the newly commissioned $S$ (1750-3500 MHz) band.   \\

One of the Large Survey Projects underway with MeerKAT is the MeerKAT International GHz Tiered Extragalactic
Exploration (MIGHTEE) survey \citep{2016mks..confE...6J}. MIGHTEE is a sensitive galaxy evolution survey over 20 deg$^2$, targeting the fields CDFS, ELAIS-S1, $XMM$-LSS and COSMOS in the $L$ and $S$ bands. $L$-band continuum maps of part of the $XMM$-LSS and COSMOS field have already been released to the public as part of the Early Science data release (ES) \citep{2022MNRAS.509.2150H}. These maps are confusion limited at a rms value of  $\sim$ 4.5 $\mu$Jy beam$^{-1}$ . Two GRGs MGTC J095959.63+024608.6 (hereafter GRG1) and MGTC J100016.84+015133.0 (hereafter GRG2) were discovered in the MIGHTEE-COSMOS field in the ES data release and presented by \citet{2021MNRAS.501.3833D}. \\

If we wish to probe the physical processes underlying GRGs and their impact on the surrounding environment, we need to measure the timescales or `ages' involved. One way to do this is by examining the radio continuum spectrum of the GRGs to determine their spectral ages. In general, the optically thin radio spectrum can be described by a power law $ S_\nu \propto \nu^{-\alpha}$ where the slope $\alpha$ is the spectral index. The spectral index of the radiation is dependent on the source of the radiation, the environment, the physical processes involved and the timescales over which they occur. At frequencies close to $\sim$1 GHz, the continuum emission is affected by synchrotron losses \citep{10.1093/mnras/stt1526}, inverse Compton scattering \citep{1999AJ....117..677B} and adiabatic losses \citep{2000AJ....119.1111B}, though the effects of adiabatic losses are not well constrained.\\

If the spectrum is taken to be dominated by synchrotron emission processes, then the lifetime $t$ of an electron is inversely proportional to its energy and the perpendicular component of the magnetic field, i.e. $t \propto (E B_\perp^2)^{-1}$. Thus, higher-energy electrons will lose their energy at a faster rate, which results in a \lq steeper' spectrum in regions of older plasma (e.g. \citealt{2000AJ....119.1111B, harwood2015spectral}) . This allows an estimate of the spectral ages of the plasma, where age is defined as the time since the plasma was last injected with newly accelerated particles. Creating a spatially resolved spectral age map of a source allows us to see in detail the dynamics and evolution of the source and how it has interacted with the environment over time.  \\

For normal sized radio galaxies, previous studies have found that $\alpha$ steepens with increased linear size and that, except at high frequencies, the rest frame spectral indices have a stronger dependence on luminosity than redshift \citep{1999AJ....117..677B,Debhade...2020b}. The spectral index is expected to steepen in the lobes away from the hotspot as electrons are accelerated at the jet termination shock. This trend is expected to extend to GRGs. Confirmation of this is an important aspect of this work, which serves as a pilot study for the spectral analysis of a larger sample of GRGs in MIGHTEE.\\

In this work, we combine MIGHTEE $L$-band data with new MeerKAT UHF-band observations to investigate the activity and history of the diffuse emission in the jets of the MIGHTEE COSMOS GRGs via the use of spectral age models. Furthermore, a newly identified third GRG found within the UHF COSMOS region is also examined. In section \ref{ch: Data} we discuss the data processing of the multi-band data and in sections \ref{ch3} and \ref{sec: Discussion} we discuss our analysis of our results. The conclusions of the analysis are summarized in section \ref{ch: conclusion}. Throughout this paper, we assume a $\Lambda$CDM cosmology of $H_0$ = 67.8 km s$^{-1}$Mpc$^{-1}$, $\Omega_\Lambda$ = 0.692 and $\Omega_M$ = 0.308 \citep{2016A&A...594A..13P}.

\section{Data}
\label{ch: Data}
For our spectral index analysis, we make use of MeerKAT observations of the COSMOS region in three frequency ranges centred on 632 MHz, 755 MHz and 1284 MHz. The 1.28 GHz data comes from $L$-band continuum maps from MIGHTEE data release 1 (DR1; \citealp{2024arXiv241104958H}).  DR1 has an expanded survey area of $\sim$ 4 deg$^2$ in the COSMOS region and an improved sensitivity (rms $\sim$ 3.5 $\muup$Jy\, beam$^{-1}$) compared to the Early Science release (1.6 deg$^2$, rms $\sim$ 4.5 $\muup$Jy\, beam$^{-1}$). Two versions of the MIGHTEE DR1 images have been released, one maximising resolution and the other maximising sensitivity. We use the lower-resolution, higher-sensitivity DR1 image in this work. Full details of the observations and data processing are given by \cite{2022MNRAS.509.2150H} and \citet{2024arXiv241104958H}. \\

The 632 MHz and 755 MHz-centred data comes from UHF band MeerKAT Open Time observations of COSMOS (ID: SCI-20210212-JD-01; PI: Delhaize), henceforth referred to as UHF-COSMOS. Details of these observations and the full-band reduction will be presented by Delhaize et al. (in prep). In this work, we only use the lower half of the UHF band (570-816 MHz).  This was to ensure the highest signal-to-noise ratio of the diffuse GRG lobe emission, since most of this emission is seen mainly in the lower part of the band. We separate the measurement set (ms) file into two equal sub-bands, dubbed UHF Low and UHF Mid respectively. This provides more spectral points to improve the accuracy of spectral index fitting.\\

The two sub-bands were separated from the original ms file using the \textsc{casa} task \textsc{mstransform} and imaged independently using the \textsc{Oxkat} pipeline\footnote{https://github.com/IanHeywood/oxkat} \citep{heywood2020oxkat} up to and including cleaning and self calibration (2GC). Since the 2GC calibration is sufficient, it is unnecessary to apply direction-dependent calibration (3GC). The primary and secondary calibrators are J0408-6545 and J1008+0730 respectively. Images are made with \textsc{wsclean} using a Briggs-weighting robust parameter of -0.5 and a pixel size of 2 arcsec to match the resolution parameters of the MIGHTEE $L$-band image as closely as possible. The images are truncated at the 30\% primary beam power level. We also apply an inner cut to the $uv$ plane at 87m. This corresponds to the shortest MeerKAT baseline at the lowest $L$-band frequency of 856 MHz to ensure that the $uv$ coverage was as similar as possible in each frequency band of our analysis.  \\

The main properties of the MIGHTEE and UHF-COSMOS data are summarised in Table \ref{tab:observations}. The $L$ and UHF band images are confusion limited, with the UHF observations reaching rms noise sensitivities of approximately 9.2 $\muup$Jy beam$^{-1}$ and 14.6 $\muup $Jy beam$^{-1}$ for UHF Low and Mid respectively. The beam size of the lower frequency UHF images are larger ($\sim 14$ arcsec) compared to the $L$-band ($\sim9$ arcsec). \\

GRG3 (see section 3.1) is entirely within the 30\% primary beam cut of the UHF-COSMOS images. However, only part of the source is within the MIGHTEE DR1 region due to the smaller field of view in the $L$-band and the conservative primary beam cut at the edges of the mosaic. We therefore re-imaged the MIGHTEE observations of a single pointing centred on 10$^\text{h}$00$^\text{m}$28.6$^\text{s}$, +2$^\circ$33$^\prime$33.8$^{\prime\prime}$ (see the appendix of \citealp{2024arXiv241104958H}) with a primary beam cut at the 30\% power level. The resulting image, which we will refer to as COSMOS$_{8}$, fully encompasses all emission associated with GRG3. However, it has a slightly poorer sensitivity (rms $\sim$ 5.1 $\muup$Jy beam$^{-1}$) compared to the full mosaic since it only contains data from a single 8-hour track.

\begin{table*}
    \centering
    \caption{Summary of the observations used for this study. The centre of the COSMOS field is 10$^\text{h}$00$^\text{m}$28.6$^\text{s}$
+02$^\circ$12$^\prime$21$^{\prime\prime}$ in the J2000 epoch and the centre of the COSMOS$_8$ field used for GRG3 in the $L$-band is  10$^\text{h}$00$^\text{m}$28.6$^\text{s}$, +2$^\circ$33$^\prime$33.8$^{\prime\prime}$. Columns: (1) Band name (2) Frequency range  (3) Average effective frequency of the observation  (5) Spatial resolution (6) Average rms noise level of the map.}
    \begin{tabular}{ccccc}
    \hline
    (1)&(2)&(3)&(4)&(5)\\
     Name           & Frequency Range      & $\nu_{\textrm{eff}}$   & Resolution    & RMS noise level    \\
                    &  (MHz)               &  (MHz)             & (arcsec$^2$)  & ($\muup$Jy beam$^{-1}$)  \\
    \hline
    \hline
     MIGHTEE $L$-band & 900-1670          & 1284             & 8.90 $\times$ 8.90    &    3.5         \\
     
     COSMOS$_8$ $L$-band & 900-1670          & 1284      & 8.69 $ \times $6.69    & 5.1               \\
     UHF Mid        &       693-816    & 755  & 11.70 $\times$ 11.70  & 9.2\\
     
     UHF Low           & 570-692        & 632         & 14.08 $\times$ 14.08               & 14.6           \\

     \hline
    \end{tabular}
    
    \label{tab:observations}
\end{table*}

\section{Results}
\label{ch3}

 \subsection{Characterising the GRGs}
As mentioned in section \ref{ch: Intro}, GRG1 and GRG2 were first identified as GRGs in the MIGHTEE Early Science images \citep{2021MNRAS.501.3833D}. These galaxies had been previously classified as extended, multiple-component radio galaxies, since the core and inner parts of the jets had been detected in previous surveys (most recently the VLA at 3 GHz; \citealt{2015A&A...574A...4V,2017A&A...602A...1S, 2019A&A...627A.142V}). However, due to brightness sensitivity limits, the most extended, diffuse and faint emission of the lobes had remained undetected until the availability of MeerKAT observations, and so these objects were not identified as giants.  GRG1 and GRG2 are both low excitation radio galaxies (LERGs) hosted by red and quenched ellipticals. For further details on GRG1 and GRG2, see \cite{2021MNRAS.501.3833D}. \\

\begin{figure*}
    \centering
    \includegraphics[width=0.9\linewidth]{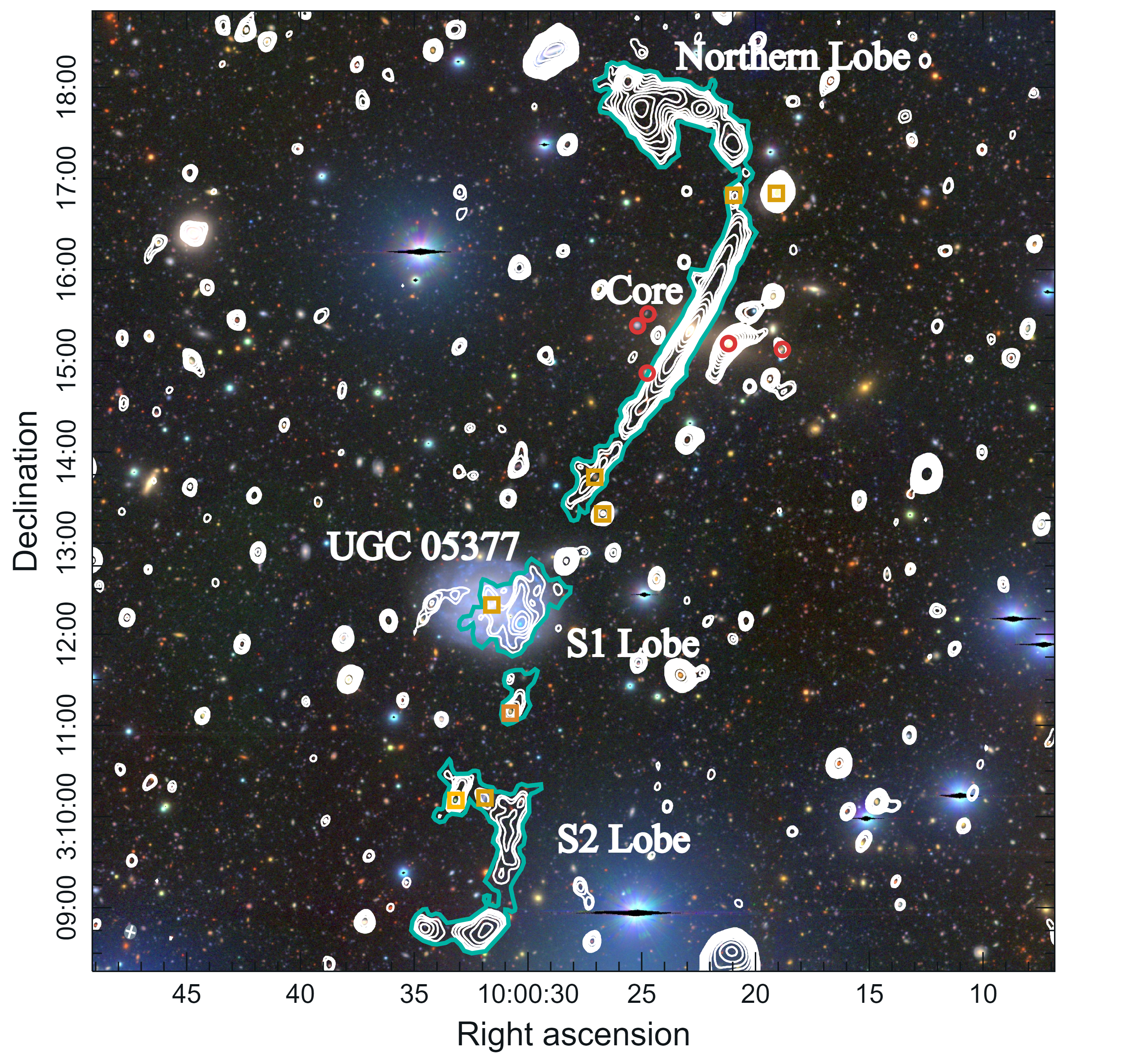}%
  \caption{The optical HSC DR2 Wide combined $g,r,i$ band image of MGTC J100022.85+031520.4 (GRG3). Overlaid in white are radio contours from the COSMOS$_8$ map, which start at 3$\sigma$ = 2.91$\times 10^{-5}$ Jy beam$^{-1}$ and increase in steps of 2$^n \sigma$ for n = {1,2,3...15}. The extent of GRG3, shown in cyan, is defined by the 3$\sigma$ contour and important features are labelled. The S1 and S2 lobes represent the upper and lower portions of the southern lobe respectively. The five nearest spectroscopically-confirmed cluster members are shown as red circles, and orange squares represent unrelated sources along the line of sight of the GRG, which have been accounted for in flux calculations.}
  \label{fig: GRG3 Optical}
\end{figure*}

We present here a third GRG in the COSMOS field for the first time. It was discovered via visual inspection in the UHF-COSMOS continuum map using a process similar to that described by \cite{prescott2018stripe}. An optical Hyper Suprime-Cam (HSC; \citealp{2018PASJ...70S...8A}) image of the host galaxy\footnote{CDS HiPS2FITS service https://alasky.cds.unistra.fr/h} is shown in Figure \ref{fig: GRG3 Optical}, with MIGHTEE $L$-band radio contours overlaid. GRG3 does not display typical Fanaroff-Riley-I/II (FRI/FRII; \citealt{1974MNRAS.167P..31F}) radio morphology. It is centre-brightened and contains a potential hot or warm spot in the northern lobe of the galaxy.  With the centre brightening, it is more likely a plumed FRI galaxy, similar to 3C 31 \citep{2008MNRAS.386..657L}. \\

The radio core is located at R.A. = 10$^\text{h}$00$^\text{m}$22.85$^\text{s}$ and Dec = +03$^\circ$15$^\prime$20.4$^{\prime\prime}$ and coincides with the known optical galaxy SDSS J100022.85+031520.4. According to the Sloan Digital Sky Survey Data Release 13 (SDSS DR13; \citealp{2017ApJS..233...25A}) the galaxy has a spectroscopic redshift of $z$ = 0.10344 $\pm$ 0.00002. Notably, it is the central galaxy and brightest cluster galaxy (BCG) in the cluster WHL J100022.9+031521 \citep{wen2012vizier}, which places it amongst 4\% of GRGs residing in clusters according to \citet{2024arXiv240308037S}. Its location at the centre of the cluster and bent morphology in the top lobe may lead it to have similar properties to wide-angle-tail (WAT) galaxies  \citep{1990AJ.....99...14B}. The host's cluster is poor, with only ten n$_{200}$ cluster members, where $n_{200}$ is the number of cluster members that occupy a region that has a density of 200 times the critical density of the universe. The total mass in the same region is $M_{200} = 1.24\times 10^{14}M_{\odot}$ (see section \ref{dynamical}). The closest cluster member is WISEA J100021.13+031511.3 (from the ALLWISE source catalogue), with a redshift of $z$ = 0.1024 and an angular separation from GRG3 of 0.45 arcminutes. \\

Several foreground galaxies overlap with the radio contours of GRG3, as seen in Figure \ref{fig: GRG3 Optical}. The northern part of the lower lobe (S1 lobe) overlaps with the star-forming galaxy UGC 05377 ($z$ = 0.007175 ± 1.33$\times 10^{-5}$; \citealt{1990ApJS...72..245S}). However, the radio contours in this region are not centred on UGC 05377, implying that at least some fraction of the emission is not due to that galaxy and is likely associated with the GRG.  We do not exclude this region from our analysis since spatially-resolved spectral modelling is performed on a pixel-by-pixel basis, so contamination by star-formation in this area will not affect our calculations along the rest of the GRG. It is assumed that the diffuse lower lobe (S2 lobe) centred at 10$^\text{h}$00$^\text{m}$32$^\text{s}$ +3$^\circ$10$^\prime$00$^{\prime\prime}$ is related to GRG3 due to the configuration of the emission. There is a potential source located at 10$^\text{h}$00$^\text{m}$33$^\text{s}$, 3$^\circ$10$^\prime$12$^{\prime\prime}$, which we assume to be a star as it is listed as such in the SDSS DR6 catalogue \citep{adelman2008sixth}, and there is no counterpart in the HSC catalogue. As with UGC 05377, we include this region in our analysis with the caveat that it might contain contaminating emission not associated with the GRG. \\

With these assumptions, the total projected angular size of the radio galaxy measured between the outer edges of the northern lobe and the S2 lobe is 12.0 arcmin. This gives it a projected linear size of 1.29 Mpc, establishing it as a GRG.  Even if the entirety of the southern lobe is discounted, the remaining extent between the northern lobe and the southern part of the inner jet has a projected length slightly above 700 kpc and the source would still be classified as a GRG. The cyan contour in Figure \ref{fig: GRG3 Optical} encompasses all emission we consider to be associated with the GRG.\\

\begin{table*}
    \centering
    \caption{Basic properties of the COSMOS GRGs studied in this work. Columns: (1) Object name (2) Right Ascension (J2000) (3)
Declination (J2000) (4) Spectroscopic redshift (5) Projected angular size (6) Projected linear size.}
    \begin{tabular}{cccccc}
    \hline
    (1) & (2) & (3) & (4) & (5) & (6) \\
     Name                               & R.A. & Dec              & $z_\textrm{spec}$  & $d$    & $D$    \\
     & (J2000) & (J2000)&  & (arcmin) & (Mpc) \\
        
    \hline
    \hline
     MGTC J095959.63+024608.6 (GRG1)   & 09$^\text{h}$59$^\text{m}$59.63$^\text{s}$ & +02$^\circ$46$^{\prime}$08.6$^{\prime\prime}$ & 0.1656      & 13.8                    & 2.42\\
     
     MGTC J100016.84+015133.0 (GRG2)    & 10$^\text{h}$00$^\text{m}$16.84$^\text{s}$ & +01$^\circ$51$^{\prime}$33.0$^{\prime\prime}$     & 0.3363      & 6.8                     & 2.04 \\

     MGTC J100022.85+031520.4 (GRG3)   & 10$^\text{h}$00$^\text{m}$22.85$^\text{s}$  & +03$^\circ$15$^{\prime}$20.4$^{\prime\prime}$  & 0.1034 & 12.0 & 1.29 \\
     \hline
     
    \end{tabular}
    
    \label{tab: GRG properties}
\end{table*}

\begin{figure*}
\centering
\begin{subfigure}{0.33\textwidth}
    \centering
  \includegraphics[width=\columnwidth]{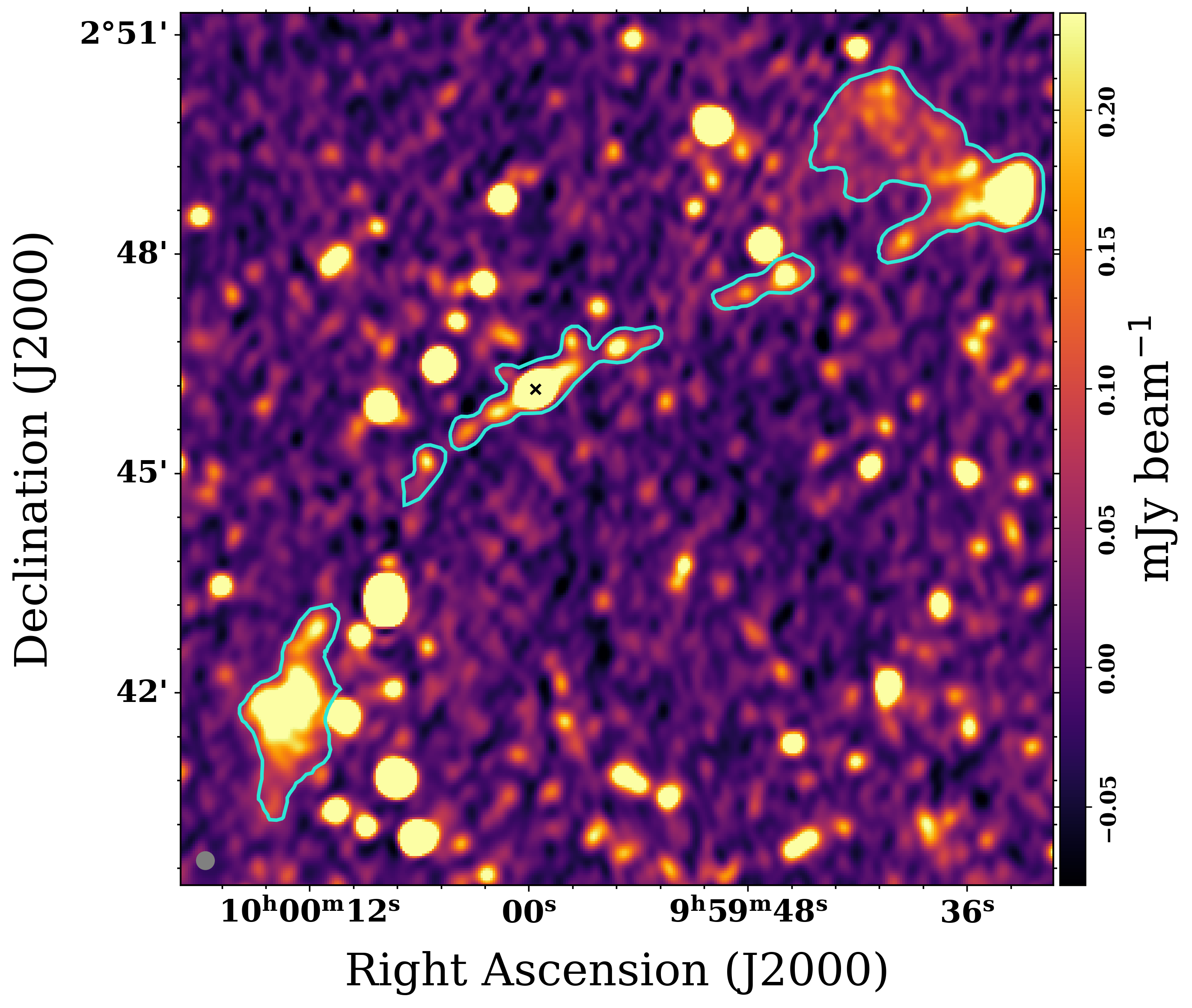}%
  \caption{UHF Low}
  \label{fig: GRG1 Low}
\end{subfigure}
\begin{subfigure}{0.33\textwidth}
    \centering
  \includegraphics[width=\columnwidth]{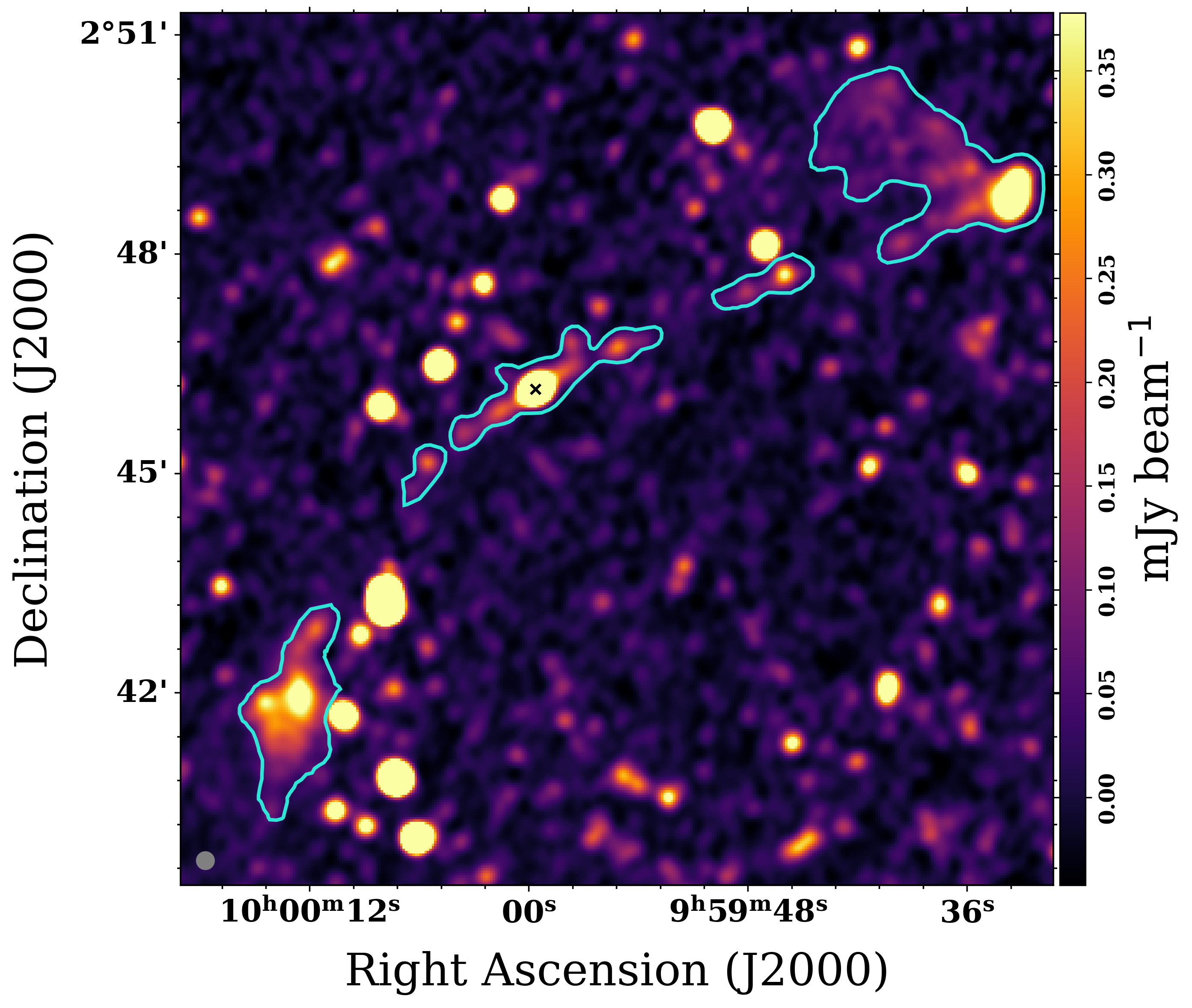}%
  \caption{UHF Mid}
  \label{fig: GRG1 Mid}
\end{subfigure}
\begin{subfigure}{0.33\textwidth}%
  \centering
  \includegraphics[width=\columnwidth]{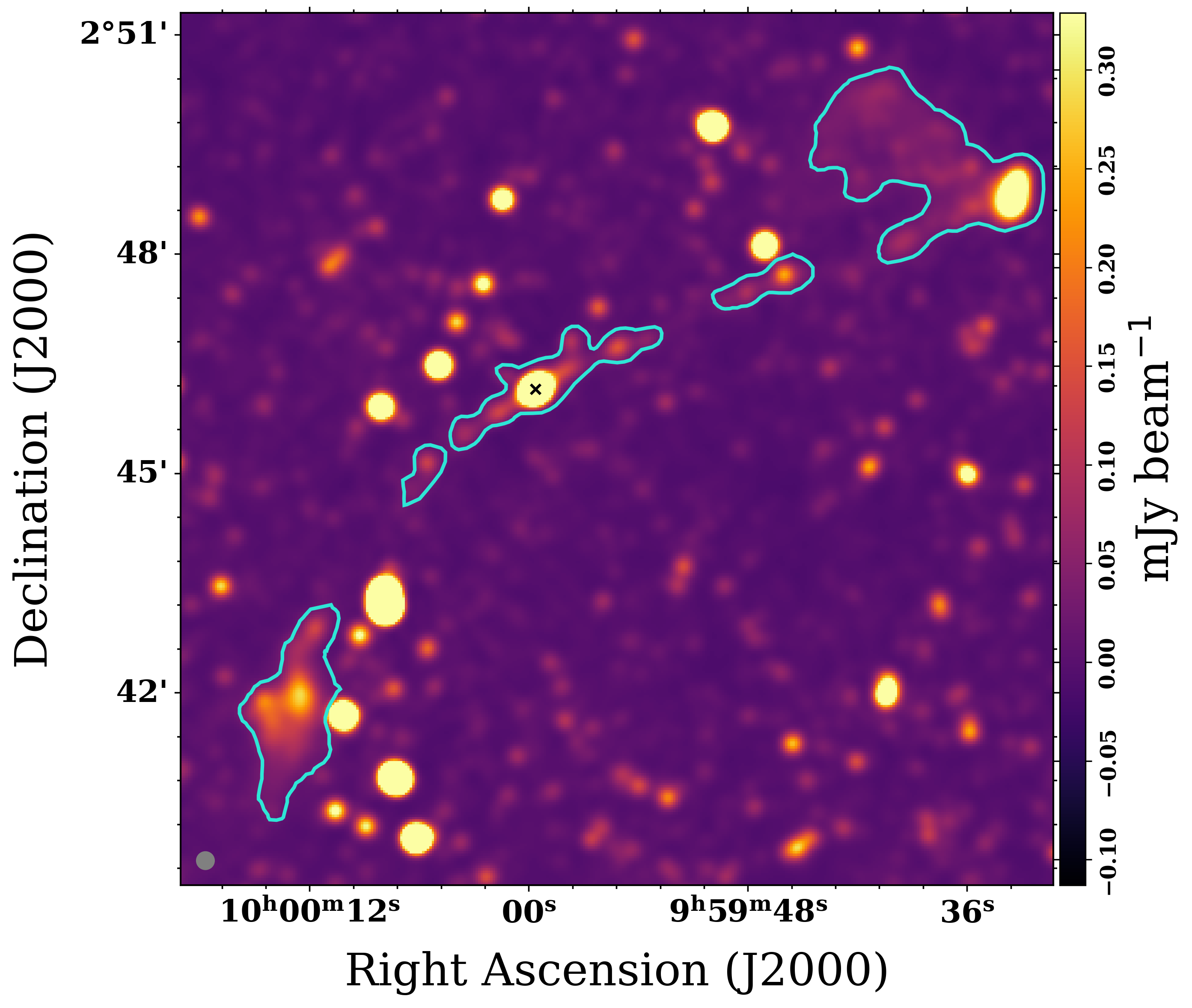}%
  \caption{$L$-band}
  \label{fig:GRG1 L}
\end{subfigure}
\caption[GRG1 in the UHF and L-bands]{MeerKAT maps of GRG1 at 632 MHz, 755 MHz and 1284 MHz, (a,b,c respectively), each smoothed to a common resolution of 14.02$\arcsec \times$ 14.02$\arcsec$. The extent of the GRG, defined by the 3$\sigma$ contour at 20.37 $\muup$Jy beam$^{-1}$ in the $L$-band, is shown in cyan.
The beam  is shown in grey in the bottom left corner of each image, and the location of the optical host AGN is shown as a black cross.}
\label{fig: GRG1}
\end{figure*}

\begin{figure*}
\centering
\begin{subfigure}{0.33\textwidth}%
  \centering
  \includegraphics[width=\linewidth]{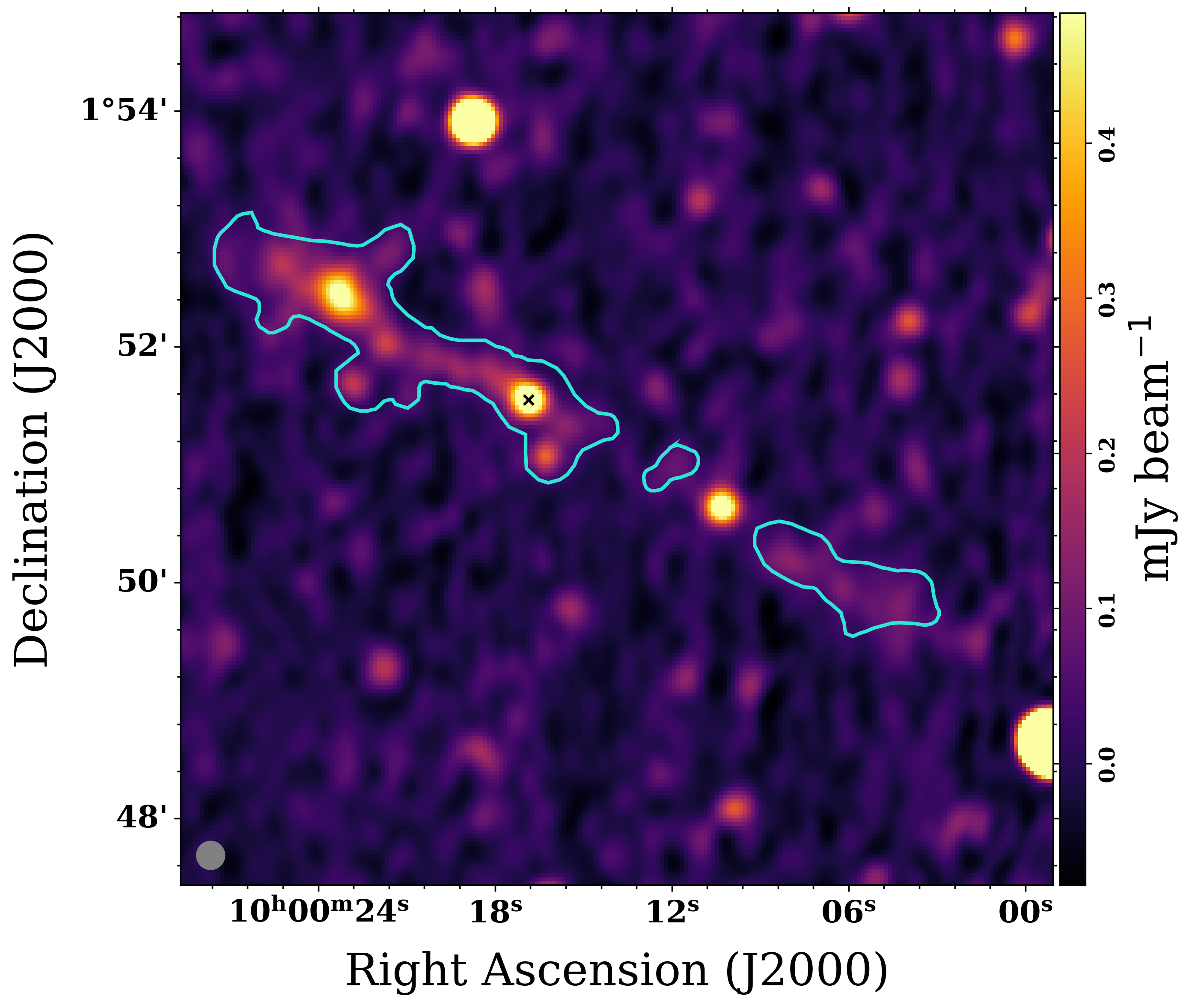}
  \caption{UHF Low}
  \label{fig: GRG2 low}
\end{subfigure}
\begin{subfigure}{0.33\textwidth}%
  \centering
  \includegraphics[width=\linewidth]{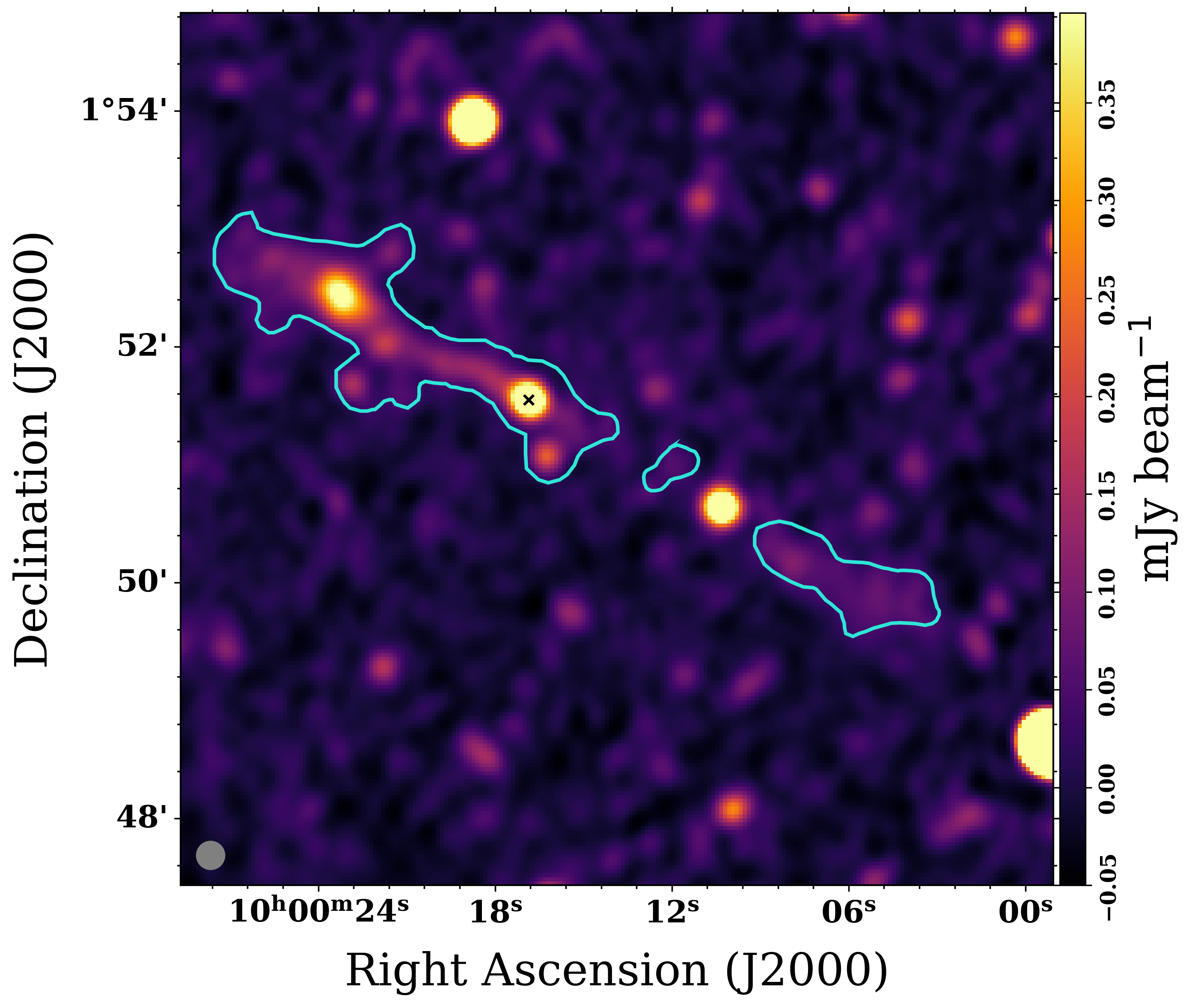}%
  \caption{UHF Mid}
  \label{fig: GRG2 mid}
\end{subfigure}
\begin{subfigure}{0.33\textwidth}%
  \centering
  \includegraphics[width=\linewidth]{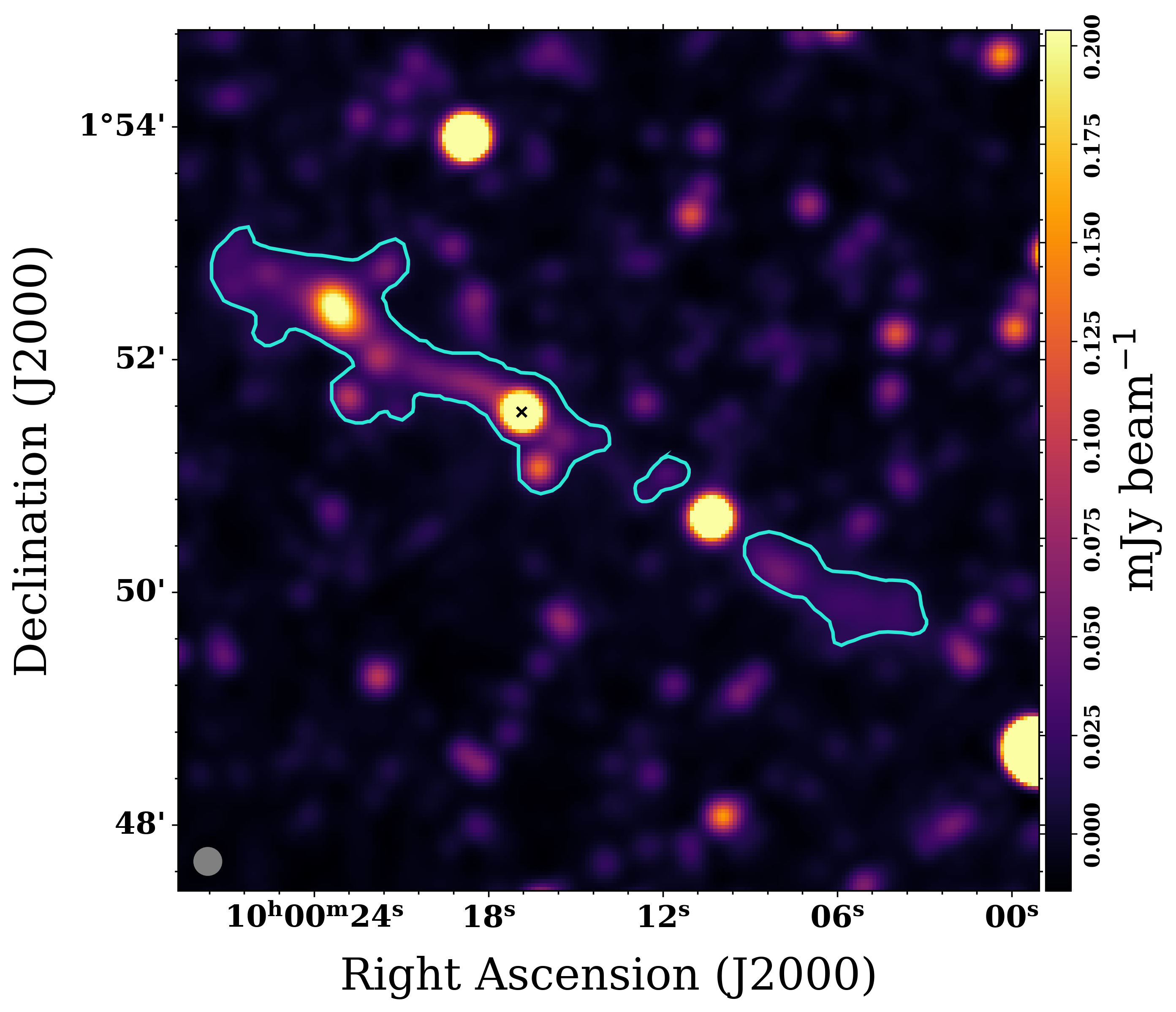}%
  \caption{$L$-band}
  \label{fig: GRG2 L}
\end{subfigure}
\caption[GRG2 in the UHF and L-bands]{As in Figure \ref{fig: GRG1} but for GRG2. The extent, marked in cyan, is taken at 13.29 $\muup$Jy beam$^{-1}$ in the $L$-band}
\label{fig: GRG2}
\end{figure*}

\begin{figure*}
\centering
\begin{subfigure}{0.33\textwidth}%
  \centering
  \includegraphics[width=\linewidth]{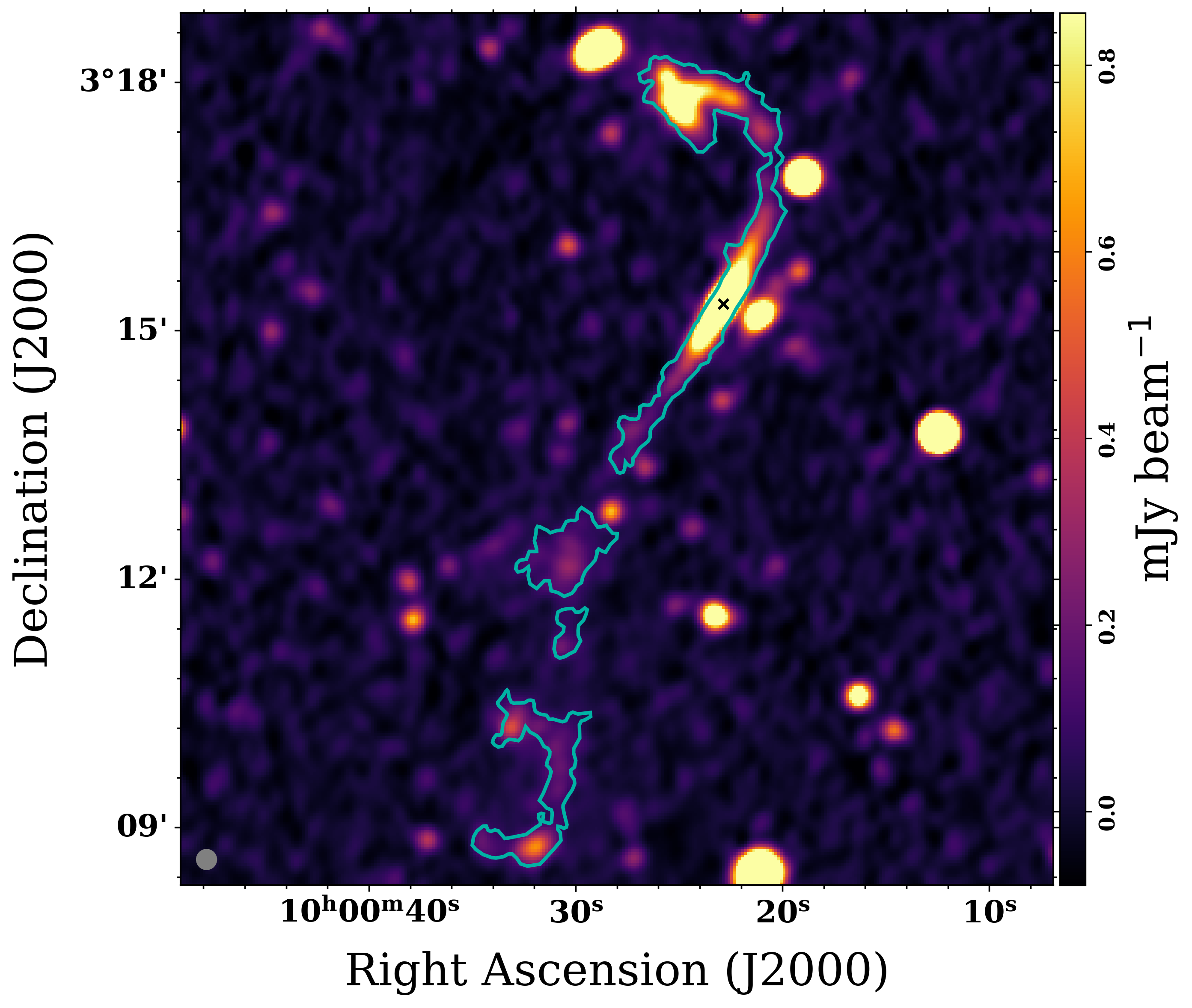}%
  \caption{UHF Low}
  \label{fig:GRG3 Low}
\end{subfigure}
\begin{subfigure}{0.33\textwidth}%
  \centering
  \includegraphics[width=\linewidth]{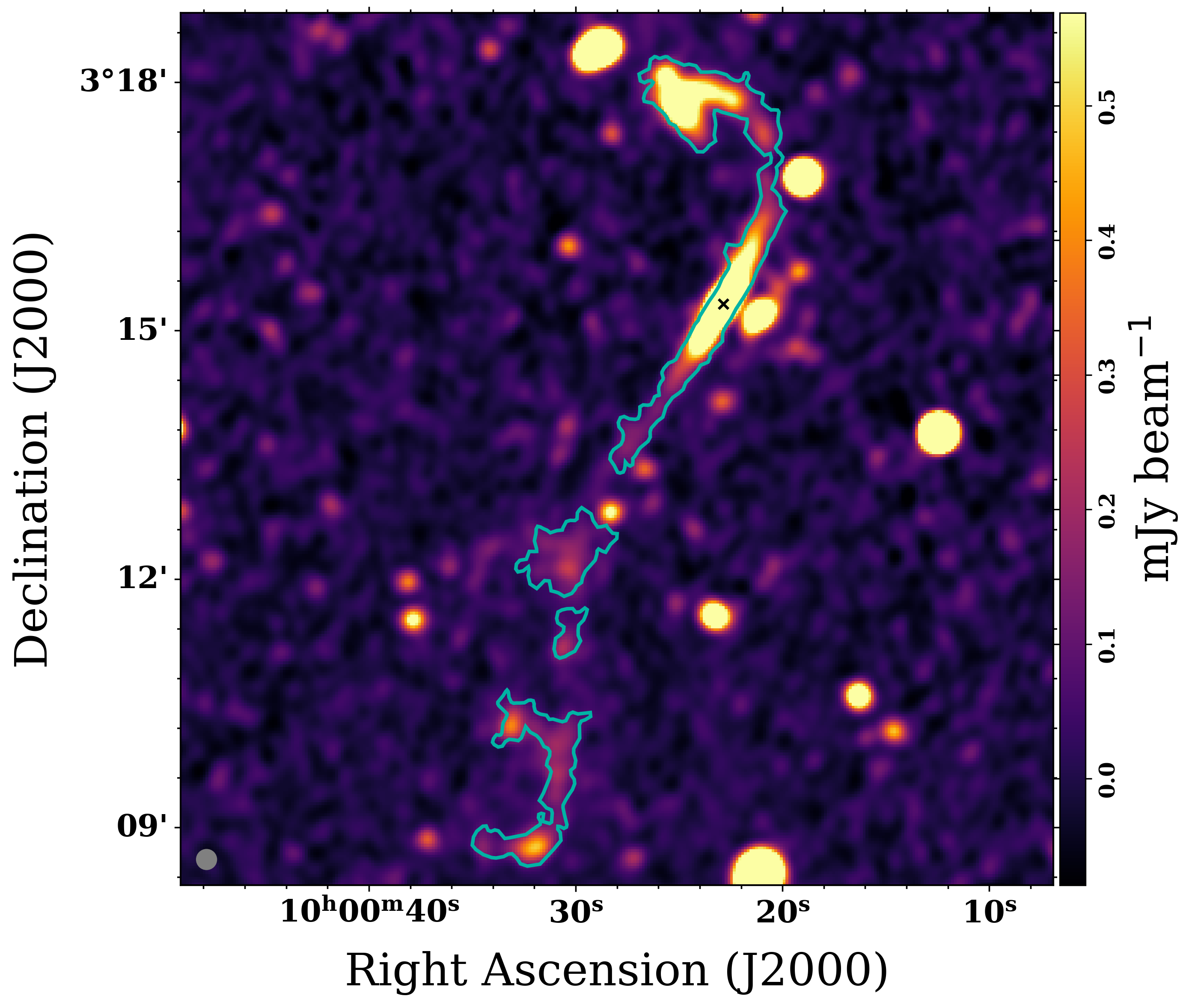}%
  \caption{UHF Mid}
  \label{fig:GRG3 mid}
  \end{subfigure}
\begin{subfigure}{0.33\textwidth}%
  \centering
  \includegraphics[width=\linewidth]{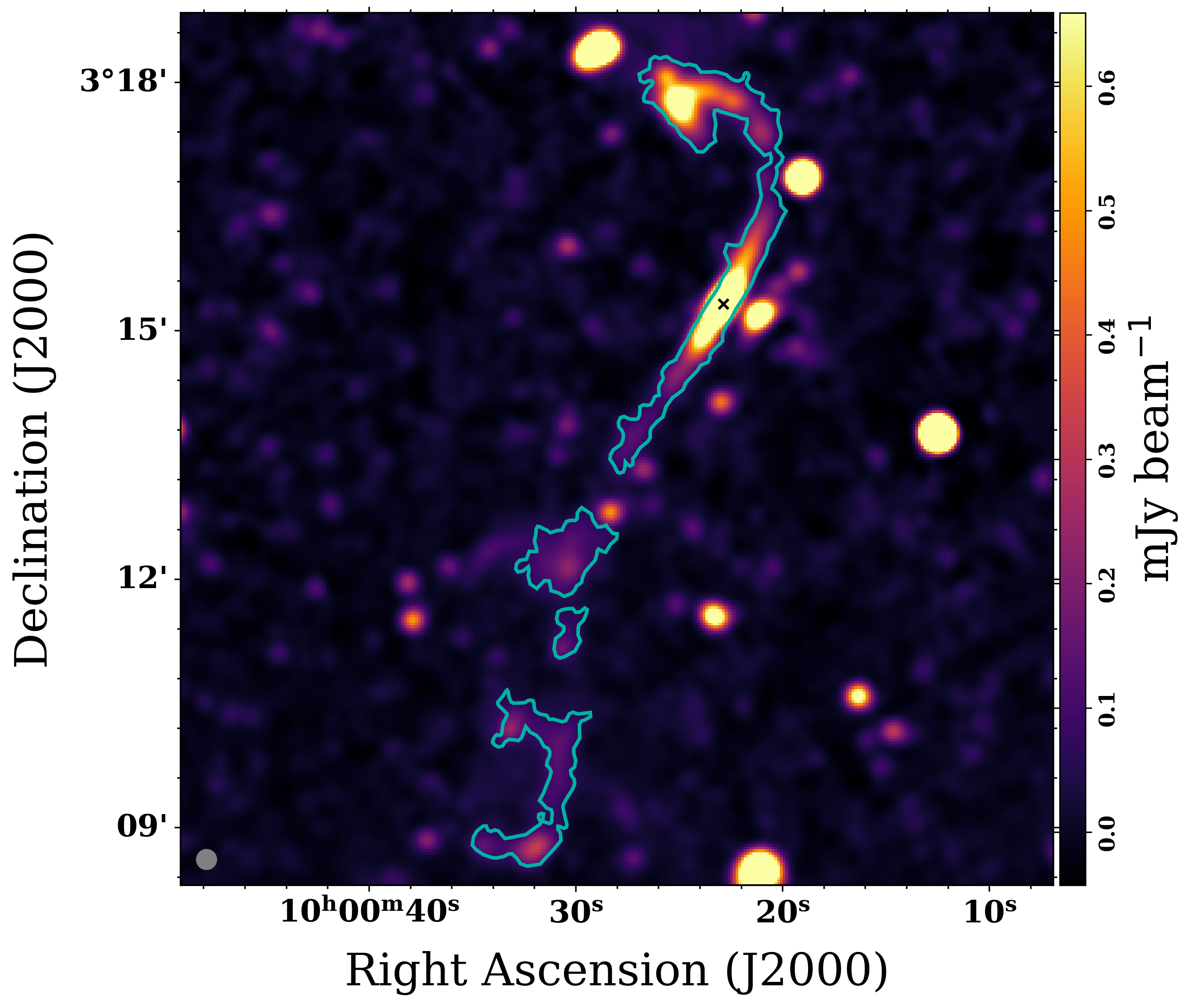}%
  \caption{$L$-band}
  \label{fig:GRG3 L-band}
\end{subfigure}
\caption[GRG3 in the UHF and $L$-bands]{As in Figure \ref{fig: GRG1} but for GRG3. The extent is defined at 29.1 $\muup$Jy beam$^{-1}$ in the $L$-band and is shown in cyan. }
\label{fig: GRG3}
\end{figure*}

The basic properties of all three GRGs found within MIGHTEE COSMOS are given in Table \ref{tab: GRG properties}. Their total intensity maps in the UHF-low, UHF-mid and $L$-bands are shown in Figures \ref{fig: GRG1}, \ref{fig: GRG2} and \ref{fig: GRG3}. The $L$-band images and associated properties differ slightly to \cite{2021MNRAS.501.3833D}, since the images have been smoothed to the same resolution as the UHF low images, and we also use the MIGHTEE DR1 rather than the ES release. The extent of each galaxy, determining its total length, is indicated by a 3$\sigma$ flux-density-contour in the $L$-band, shown in cyan. The $L$-band image is used in preference to the UHF images to determine the full extent due to its better sensitivity revealing more of the diffuse outer lobe emission. \\ 

Table \ref{tab: Flux properties} presents the integrated flux densities ($S_{\textrm{int}}$), integrated spectral indices ($\alpha$) and total radio power of each GRG from the intensity maps within the defined extent. The total integrated flux density is the sum of the northern lobe, southern lobe and core contributions to the flux density. The uncertainty of $S_{\textrm{int}}$ is dominated by the $\sim$ 10 percent uncertainty on the flux scale; however, we also combine it in quadrature with the thermal and confusion noise. A thermal noise map was created using \textsc{Breizorro}\footnote{https://github.com/ratt-ru/breizorro}, with a threshold of 3$\sigma$. A confusion noise map was created with the Python Blob Detector and Source Finder (\textsc{PyBDSF}; \citealt{2015ascl.soft02007M}) using a pixel threshold of 5$\sigma$ and an island threshold of 2$\sigma$. For each map, the corresponding contribution to the total flux density uncertainty was taken to be the average value of all pixels encompassed by the GRG emission (i.e. cyan lines in Figures \ref{fig: GRG1},\ref{fig: GRG2} and \ref{fig: GRG3}). \\

The frequency attributed to each flux measurement is taken as that of the central spectral channel covered by the data. The exception to this is the $L$-band DR1 data where the average effective frequency of the map has been accurately determined to be 1.284\, GHz (see \citet{2024arXiv241104958H} for details). The effective frequencies at the positions of GRG1 ($1.289 \pm 0.002 $ GHz) and GRG2 ($1.315 \pm 0.001$ GHz) are very similar to the average, and so we choose to use the average value of 1.284 GHz as the frequency of the $L$-band data throughout this work. This is for consistency with the measurements in the UHF sub-bands and in the COSMOS$_8$ $L$-band (for GRG3), where more detailed effective frequency maps are not available. We have verified that using, for example, a frequency of 1.31 GHz for GRG2, does not change our results. \\

From Table \ref{tab: Flux properties}, we see that GRG3 has a significantly lower radio power than GRG1 and GRG2 at all frequencies. \\

\begin{table*}
\caption{Integrated flux density and radio power of the GRGs in each band as well as the integrated spectral index values. Columns: (1) Object name (2,3,4) Integrated flux density at 632 MHz, 755 MHz and 1.28 GHz respectively (5) Integrated spectral index calculated from the integrated flux density at 632 MHz and 1284 MHz (6,7,8) Power at an effective frequency at 632 MHz, 755 MHz and 1.28 GHz respectively, determined using the spectral index in column 5.}
    \centering
    \begin{tabular}{cccccccc}
    \hline
    (1) & (2) & (3) & (4) & (5) & (6) & (7) & (8)  \\
    Name   & $S_{\textrm{int}, 632\textrm{MHz}}$ &  $S_{\textrm{int}, 755\textrm{MHz}}$ & $S_{\textrm{int}, 1284\textrm{MHz}}$  & $\alpha_{632\textrm{MHz}}^{1284\textrm{MHz}}$ & $P_{632 \textrm{MHz}}$ &  $P_{755\textrm{MHz}}$ & $P_{1284 \textrm{MHz}}$  \\
             & (mJy) & (mJy) & (mJy) & &   (10$^{24}$ W Hz$^{-1}$) & (10$^{24}$ W Hz$^{-1}$)  &(10$^{24}$ W Hz$^{-1}$) \\
    \hline
    \hline
    GRG1 & 33.20 $\pm$ 2.00 & 29.10 $\pm$ 1.70 &  19.10 $\pm$ 1.10 & 0.78 $\pm$ 0.12 & 2.590 $\pm$ 0.270 & 2.230 $\pm$ 0.230 & 1.490 $\pm$ 0.160 \\

    GRG2  & 7.47 $\pm$ 0.53 & 5.98 $\pm$ 0.45 &  3.88 $\pm$ 0.25 & 0.92 $\pm$ 0.13 & 3.000 $\pm$ 0.350 & 2.090 $\pm$ 0.240 & 1.380 $\pm$ 0.160 \\
    
    GRG3  & 32.30 $\pm$ 2.00 & 28.70 $\pm$ 1.80 &  21.20 $\pm$ 1.30 &  0.59 $\pm$ 0.12 & 0.890 $\pm$ 0.092 & 0.802 $\pm$ 0.082 & 0.597 $\pm$ 0.061 \\
    \hline
    \end{tabular}

\label{tab: Flux properties}
\end{table*}

\subsection{Spectral indices and ages}
We use Broadband Radio Astronomy Tools (\textsc{brats}) to estimate the spectral indices and spectral ages of the three GRGs on a pixel-by-pixel basis \citep{10.1093/mnras/stt1526,harwood2015spectral}. \textsc{Brats} takes flux density maps at different frequencies and calculates the spectral index of each pixel using a least squares regression in log space. \textsc{brats} can then convert spectral indices to spectral ages given various input parameters, such as the magnetic field strength ($B$).

\subsubsection{Spectral index maps}
\begin{figure*}
    \centering
    \begin{subfigure}{0.32\textwidth}
    \includegraphics[width=\columnwidth]{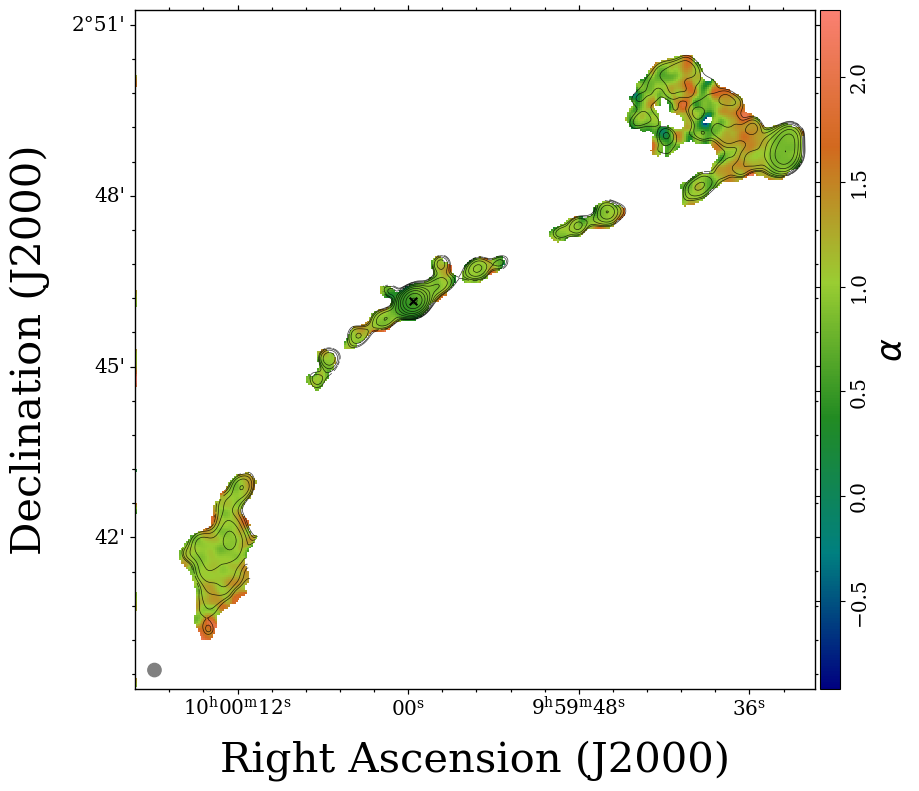}
    \caption[GRG1 spectral index map]{GRG1}
    \label{fig:specindex GRG1}
    \end{subfigure}
    \begin{subfigure}{0.32\textwidth}
    \includegraphics[width=\textwidth]{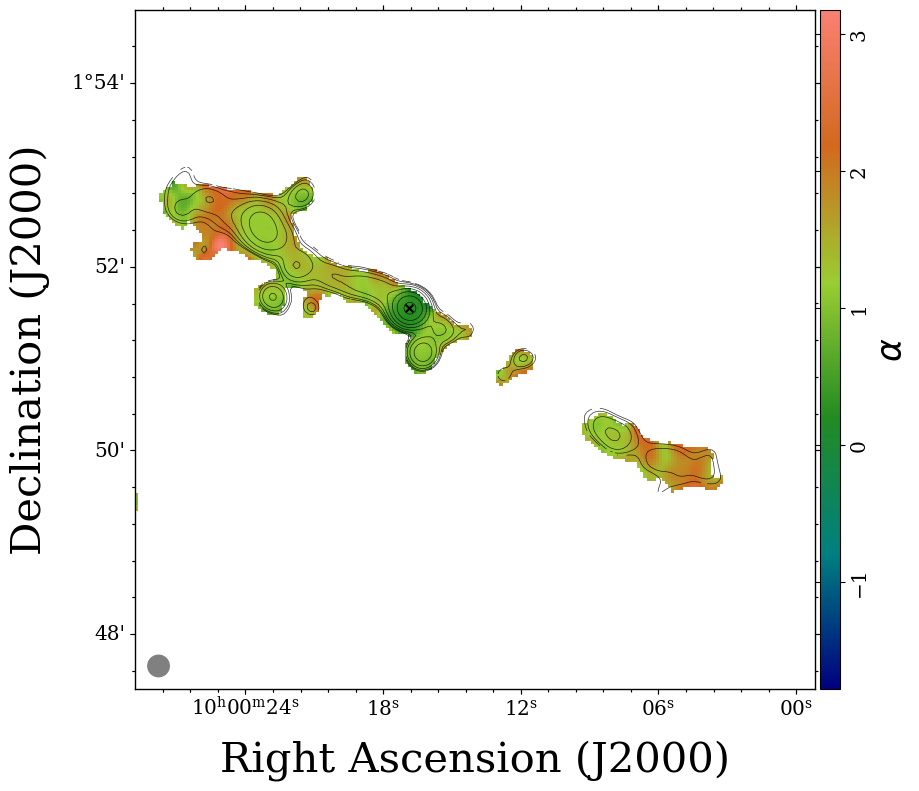}
    \caption[GRG2 spectral index map ]{GRG2 }
    \label{fig:specindex GRG2}
    \end{subfigure}
    \begin{subfigure}{0.32\textwidth}
        \includegraphics[width= \columnwidth]{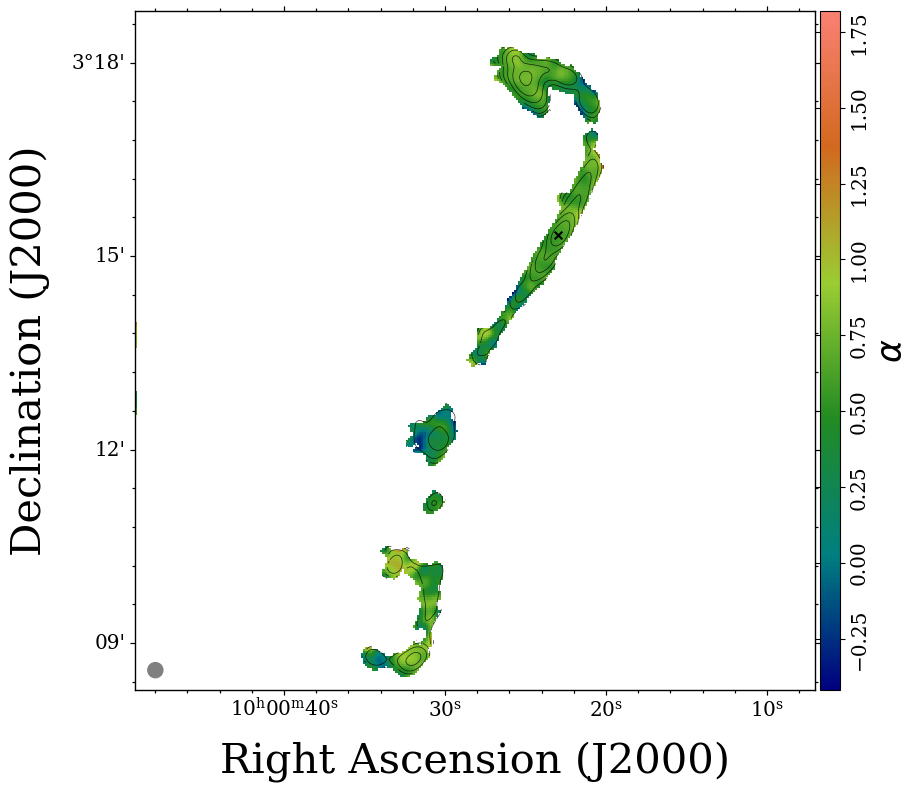}
    \caption[GRG3 spectral index map]{GRG3}
    \label{fig:specindex GRG3}
    \end{subfigure}
    \caption{Spectral index maps of the GRGs as created in \textsc{brats}.  Figure \ref{fig:specindex GRG1} has estimated spectral indices within the range of $ -0.58 \leq \alpha \leq 2.14$. Figure \ref{fig:specindex GRG2} contains values ranging from $ -0.06 \leq \alpha \leq 3.18$. Figure \ref{fig:specindex GRG3}  has a spectral index range of $ -0.42 \leq \alpha \leq 1.33$. The colourbars show the full range of indices in each case. Shown in black are the overlaid contours of the corresponding $L$ map, starting at 3$\sigma$ and increasing with intervals of 2$^n\sigma$ where n = 0,1...15. The beam is indicated in grey in the bottom left corner.}
    \label{fig: specindex}
\end{figure*}

We produce spectral index maps with \textsc{brats} using maps at the three frequencies 632 MHz, 755 MHz and 1284 MHz. The resultant maps are shown in Figure \ref{fig: specindex}. GRG2 displays a spectral index map typical of an FR II, with a flat core ($\alpha \sim 0.3$) and gradual steepening along the jets towards the outer end of the lobes ($\alpha > 2$). GRG1 displays a steeper core ($\alpha \sim 0.6$) than what we expect, though it is still shallow relative to its lobes ($ \alpha<1.6$). This indicates a possible restarted jet, which would be direct evidence of the duty cycle. GRG3 however, is relatively uniform in comparison. The majority of pixels have values in the range of  $ 0.4 <\alpha < 0.7$. We do not see any resolved core, and the inner jets have an average of $ \alpha \sim 0.6$.

\subsubsection{Magnetic Fields}
To calculate spectral ages, \textsc{brats} requires information on the magnetic field strength $B$ in the GRG lobes. We estimate $B$ using  \textsc{pysynch}\footnote{https://github.com/mhardcastle/pysynch} \citep{1998MNRAS.294..615H}, which models the synchrotron radiation and inverse Compton emission of radio galaxies assuming equipartition, i.e., that the total energy densities of cosmic ray electrons are equal to that of the magnetic field \citep{https://doi.org/10.1002/asna.200510366}. This assumption is made on the basis that cosmic ray particles and magnetic fields are strongly coupled and exchange energy until equilibrium is reached. \textsc{Pysynch} approximates the radio lobe morphologies as ellipses and the radio continuum spectrum as a power law. For each GRG lobe, we find the total flux density within an ellipse drawn from the core to the end of the lobe. We then take the average of the semi-major axes, semi-minor axes, and flux densities of the two lobes for each GRG, and provide this to \textsc{pysynch}. These input parameters are presented in Table \ref{tab: equipartition}. \\

The equipartition fields $B_{\textrm{eq}}$ are 1.00 $\muup$G, 0.95 $\muup$G  and 1.50 $\muup$G for GRGs 1, 2 and 3 respectively, giving an average equipartition magnetic field
of $B_{\textrm{eq}}$ = 1.15 ± 0.02 $\muup$G. This is just in the range of $B_{\textrm{eq}} \sim$ 1-16 $\muup$G of GRGs reported by \cite{2023JApA...44...13D} in their review of GRG properties. However, recent observations have shown that the true magnetic field lies closer to 40 percent of the equipartition value \citep{2005ApJ...626..733C,2017MNRAS.467.1586I, 10.1093/mnras/stz3396}. We therefore assume the magnetic fields of the GRGs are 0.4$B_{\textrm{eq}}$, where $B_{\textrm{eq}}$ is the equipartition magnetic field strength. The final magnetic fields $B$ = 0.4$B_{\textrm{eq}}$ are also shown in Table \ref{tab: equipartition}. \\

\begin{table}
    \centering
    \caption{Input parameters for the \textsc{pysynch} code to calculate the equipartition energy and the final magnetic fields. Columns: (1) Name of the source, (2) The semi-major axis of the ellipse used to estimate the lobe size, (3) The semi-minor axis of the ellipse, (4) Flux density as dictated by the ellipse, (5) The final magnetic field, which is taken to be 0.4$B_{\textrm{eq}}$,  where $B_{\textrm{eq}}$ is the equipartition field given by \textsc{pysynch}.  }
    \begin{tabular}{ccccc}
    \hline
    (1) & (2) & (3) &  (4) & (5) \\
      Name  & $a$ & $b$ & $S_\nu$ & $B$\\
            & (arcsec)  & (arcsec) & (mJy) & ($\times 10^{-1}\muup$G) \\
            \hline
            \hline
    GRG1    & 90    & 64    & 9.79 & 4.18 \\
    
    GRG2    & 63    & 45    & 2.56  & 3.80 \\
    
    GRG3    & 60    & 45    & 8.65  & 5.87 \\
    \hline
    \end{tabular}
    
    \label{tab: equipartition}
\end{table}

\subsubsection{Spectral ages}
The spectral age maps for the three GRGs are determined with the maps at 632 MHz, 755 MHz and 1284 MHz using two of the single injection models within \textsc{brats}: the standard Jaffe-Perola model (JP; \citealt{1973A&A....26..423J}) and JP-Tribble \citep{1993MNRAS.261...57T} model. Both models assume that the initial electron energy distribution is described as a power law $N(E) = N_0 E^{-2\alpha_{\textrm{inj}} -1}$, where $2\alpha_{\textrm{inj}} +1$ is the power law index of the initial injected energy distribution, and assume the losses for the distribution as synchrotron self-absorption and inverse Compton losses. Both models need the redshift of the galaxy, the magnetic field and the injection index ($\alpha_{\textrm{inj}}$). The models differ in that the standard JP model applies a constant magnetic field, while the Tribble method varies the magnetic field in space by applying a Gaussian random field to the source, so that the electrons diffuse over spatial regions of various magnetic field strengths. \textsc{brats} accounts for magnetic field losses due to the cosmic microwave background in both models. The Tribble method has been shown to be more physically plausible than the JP method (\citealt{10.1093/mnras/stt1526,harwood2015spectral}), however it is much more computationally expensive. While an analysis of both was feasible for the small sample of three GRGs, ensuring that the standard JP method is in agreement with the Tribble estimates would be beneficial for future studies. For a more in-depth explanation of the two methods, see \citet{harwood2015spectral,10.1093/mnras/stt1526}. \\

For the spectral age fitting, a circular background region of 30 arcsec in diameter was used to calculate the flux density image rms noise. The spectral age of a given pixel is set to only be calculated if there is a flux density above 1$\sigma$ in every input frequency, due to the UHF sub-bands having lower S/N ratios than the $L$-band image. The flux calibration uncertainty is set to be 10 percent, which is the standard for \textsc{brats} calculations. The age resolution is set as 10 Myr with a 5-level iteration, with a maximum age of 150 Myr and a minimum of 0 Myr as reasonable bounds.\\

Since the injection index of GRGs has yet to be observationally constrained, we adopt a constant injection index of $\alpha_{\textrm{inj}}$ = 0.5. This is chosen in accordance with \cite{1978MNRAS.182..147B} and \cite{carilli1991multifrequency} as the lowest value that can be theoretically observed and a common value used in the literature. Since recent studies \citep{10.1093/mnras/stt1526} have shown that injection indices can be as high as $\sim$ 0.8 \textcolor{red}, a preliminary estimate of the injection index was carried out with \textsc{brats}. The three GRGs in this study are found to have values between $0.5< \alpha_{\textrm{inj}}< 0.65 $, ensuring that $\alpha_{\textrm{inj}}$ = 0.5 is an appropriate choice. We note that if the chosen injection index is artificially low, it will artificially increase the absolute values of the calculated spectral ages; however, it will not affect the \textit{relative} distribution of ages throughout a source.  \\

\begin{figure*}
\centering
\begin{subfigure}{\columnwidth}%
  \centering
  \includegraphics[width=\linewidth]{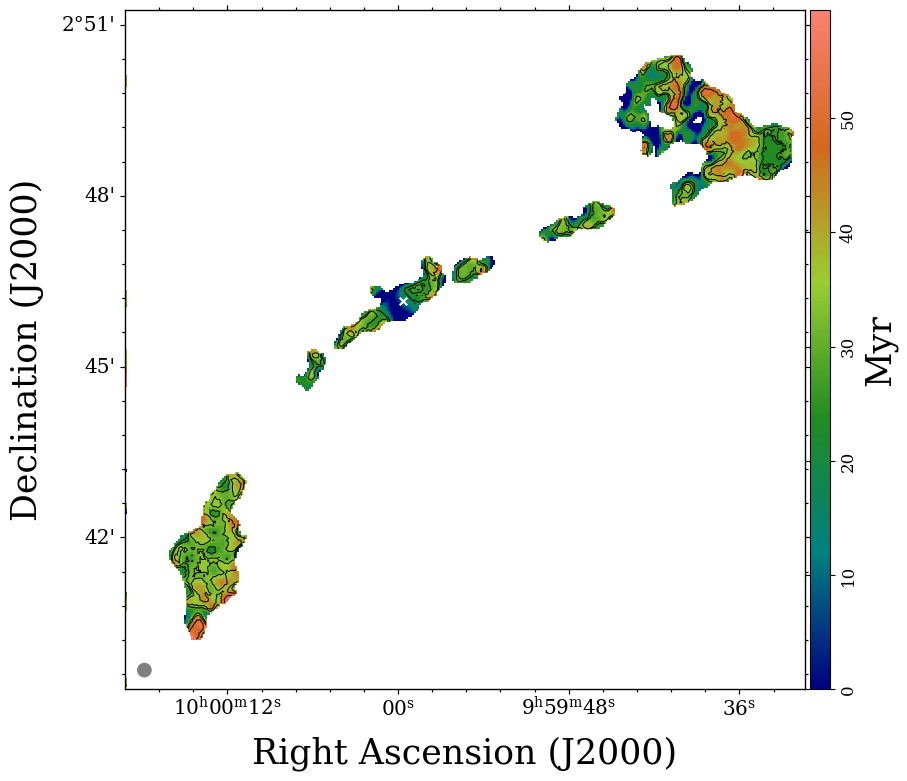}%
  \caption{Jaffe-Perola Standard}
  \label{fig:GRG1 JP}
\end{subfigure}
\begin{subfigure}{\columnwidth}%
  \centering
  \includegraphics[width=\linewidth]{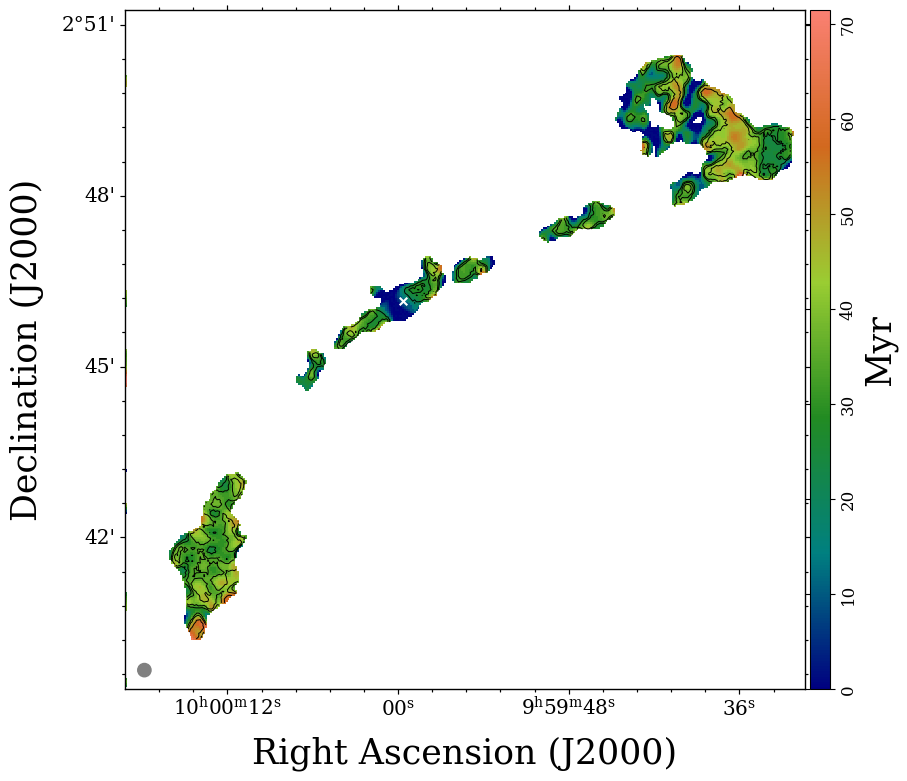}%
  \caption{Jaffe-Perola Tribble}
  \label{fig:GRG1 Tribble}
\end{subfigure}
\caption[GRG1 spectral ages]{Spectral ages for GRG1 as determined by \textsc{brats} using different fitting models with $L$-band contours overlaid (see Figures \ref{fig: GRG1} and \ref{fig:specindex GRG1} for details). The models were run using an injection index of 0.5 and a magnetic field of 0.418 $\muup$G. The maximum JP age is 60 $\pm$ 11 Myr and the maximum Tribble age is 68 $\pm$ 13 Myr. Percentage uncertainty contours are shown in black at 20\%, 30\% and 40\%.} 
\label{fig:GRG1 specages}
\end{figure*}

\begin{figure*}
\centering
\begin{subfigure}{\columnwidth}%
  \centering
  \includegraphics[width=\linewidth]{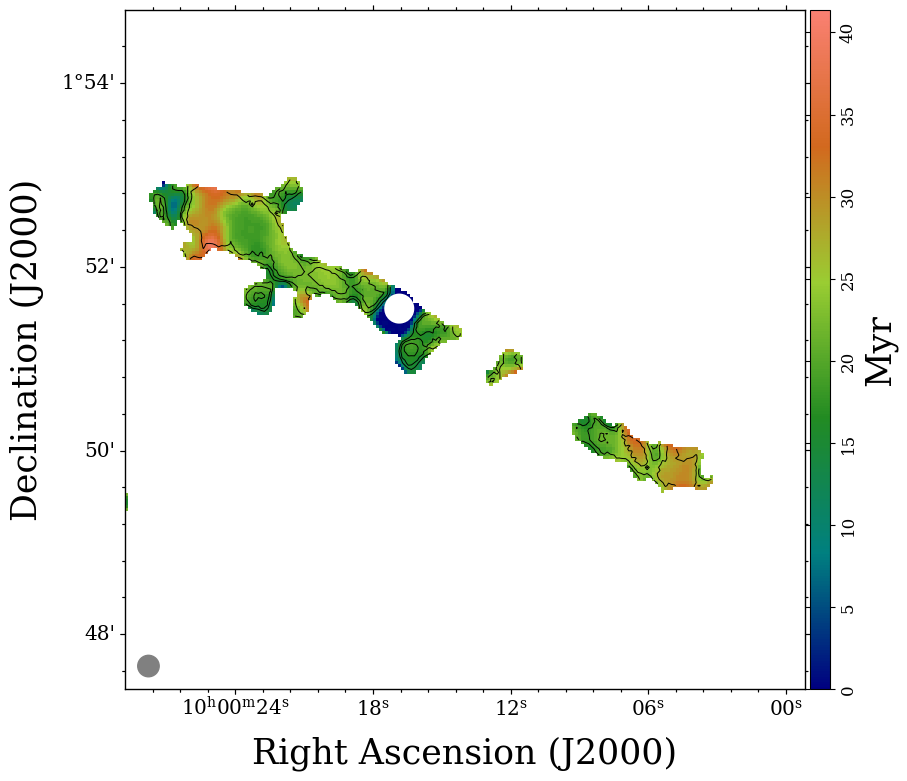}%
  \caption{Jaffe-Perola Standard}
  \label{fig:GRG2 JP}
\end{subfigure}
\begin{subfigure}{\columnwidth}%
  \centering
  \includegraphics[width=\linewidth]{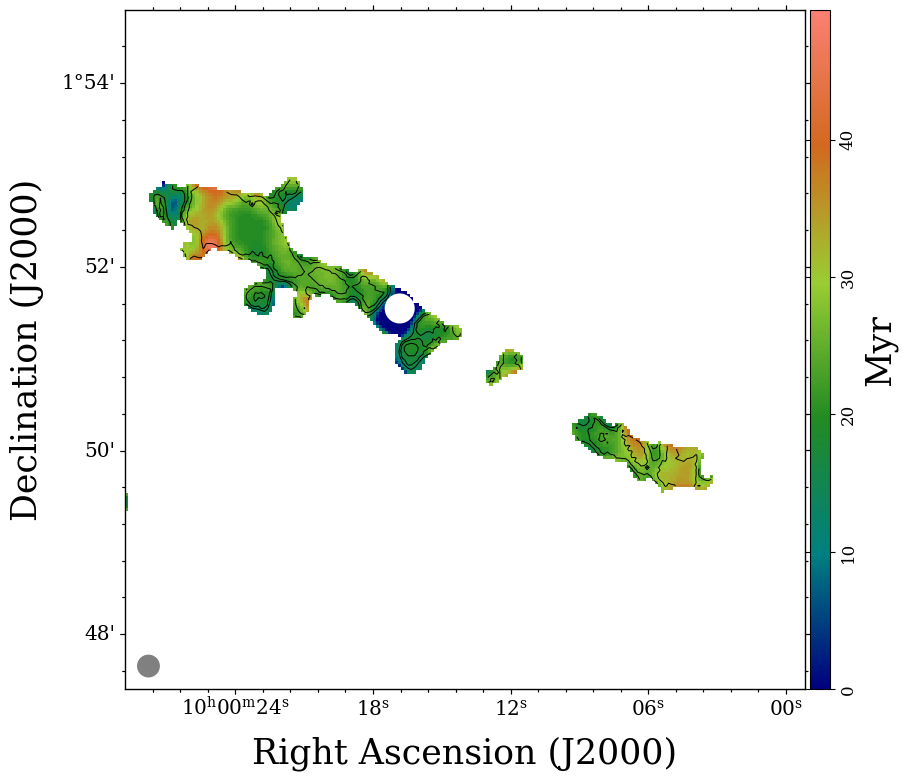}%
  \caption{Jaffe-Perola Tribble}
  \label{fig:GRG2 Tribble}
\end{subfigure}
\caption[GRG2 spectral ages]{As for Figure \ref{fig:GRG1 specages} but for GRG2. The magnetic field is 0.380 $\muup$G, and the resulting maximum ages are 39.2 $\pm$ 7.5 Myr and 47.0 $\pm$ 11.0 Myr according to the JP and JP Tribble models respectively. Percentage uncertainty contours are shown in black at 20\%, 30\% and 40\%.}
\label{fig:GRG2 specages}
\end{figure*}

\begin{figure*}
\centering
\begin{subfigure}{\columnwidth}%
  \centering
  \includegraphics[width=\linewidth]{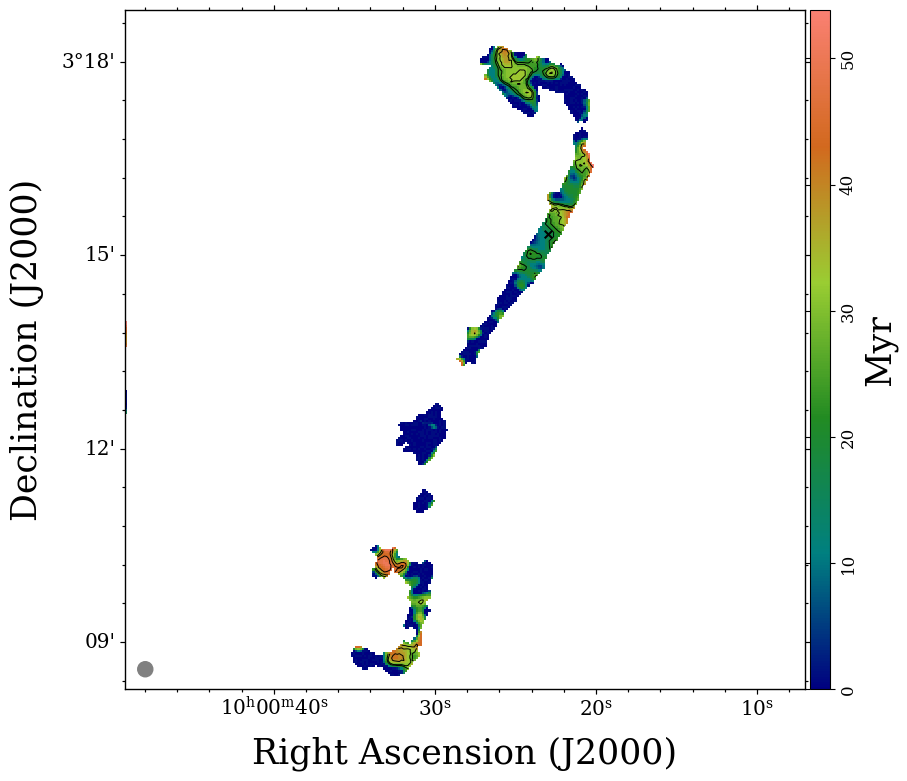}%
  \caption{Jaffe-Perola Standard}
  \label{fig:GRG3 JP}
\end{subfigure}
\begin{subfigure}{\columnwidth}%
  \centering
  \includegraphics[width=\linewidth]{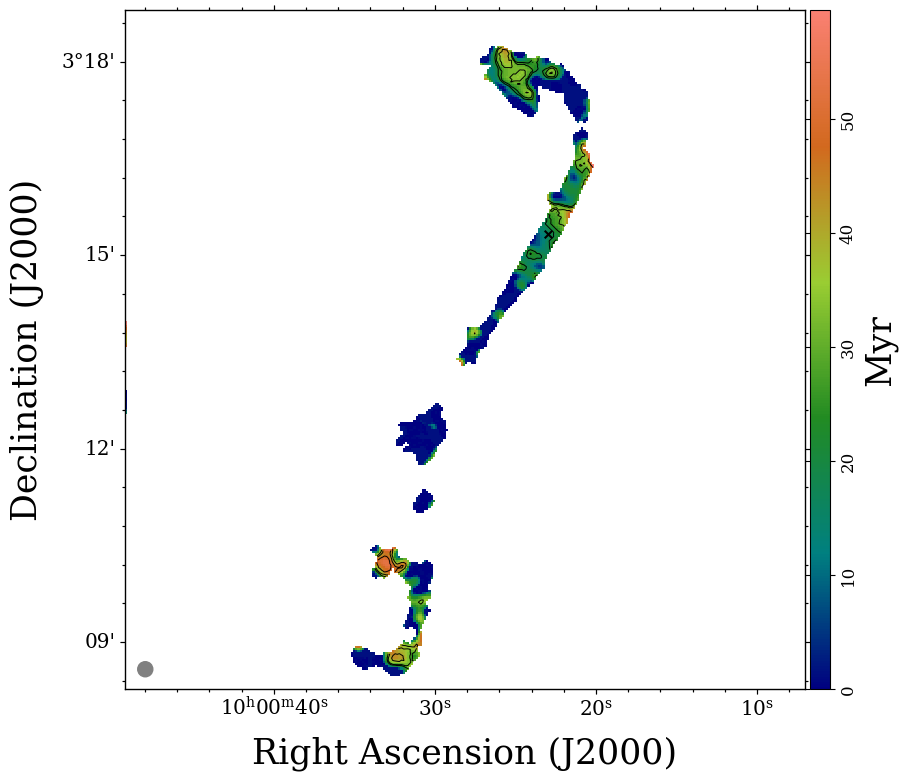}%
  \caption{Jaffe-Perola Tribble}
  \label{fig:GRG3 Tribble}
\end{subfigure}
\caption[GRG3 spectral ages]{As for Figure \ref{fig:GRG1 specages} but for GRG3. The magnetic field is 0.587 $\muup$G, and the resulting maximum ages are 64 $\pm$ 31 Myr and 67 $\pm$ 36 Myr according to the JP and JP Tribble models respectively. Percentage uncertainty contours are shown in black at 40\%, 60\% and 80\%.} 
\label{fig:GRG3 specages}
\end{figure*}

The resultant spectral age maps as calculated by both the Jaffe-Perola standard and Tribble models are shown in Figures \ref{fig:GRG1 specages}, \ref{fig:GRG2 specages} and \ref{fig:GRG3 specages}. The average ages of each component are given in Table \ref{tab:Specages JP}. Uncertainties for the spectral ages are taken from uncertainty maps produced by \textsc{brats} for each model. \textsc{brats} also calculates spectral age uncertainties on a pixel-by-pixel basis. In Figures \ref{fig:GRG1 specages} and \ref{fig:GRG2 specages}, we overlay contours connecting pixels with spectral age uncertainties of 20, 30 and 40 per cent for GRG1 and GRG2. The contours in Figure \ref{fig:GRG3 specages} represent the 40, 60 and 80 per cent uncertainties for GRG3. Possible reasons for these larger spectral age uncertainties are discussed in Section \ref{sec: Discussion}.  Both models give a similar distribution of spectral ages to each other and to their respective spectral index maps which is expected, and suggests that in future studies, the JP method may be adequate for spectral age analyses. The Tribble method gives slightly higher age estimates (with the difference in values being more pronounced at higher ages). However, we can see from Table \ref{tab:Specages JP} that the values agree with one another within their uncertainty budgets. For all three GRGs, the youngest ages are seen in the centre and in the hotspots of the GRG, with older electrons in the lobes surrounding the hotspots as well as in the jets, as expected. \\

\begin{table}
\caption{Measured average spectral ages within the core and lobes of the GRGs. Columns: (1) The source, (2) The model of spectral ageing, (3,4,5) Average spectral age in the core or inner jets in the case of GRG3, the nothern and southern lobes respectively. }
    \centering
    \begin{tabular}{ccccc}
    \hline
    (1)&(2)&(3)&(4)&(5)\\
     Source & Model           & Core/inner jets   & Northern Lobe   & Southern Lobe    \\
             &       & (Myr)     & (Myr)     & (Myr)       \\   
    \hline
    \hline
    & JP   &$2^{+15.0}_{-2.8}$ &$35.2^{+7.2}_{-9.3}$ & $32.9^{+6.8}_{-8.3}$ \\
    GRG1 & & & & \\

    & Tribble  &$3^{+14.0}_{-2.1}$ &$37.6^{+7.5}_{-9.5}$ & $34.5^{+7.9}_{-9.0}$ \\
    \hline

    & JP   &- &$22.7^{+3.5}_{-4.2}$ & $25.7^{+5.0}_{-6.3}$ \\

    GRG2 & & & & \\
     & Tribble   & - &$24.7^{+3.4}_{-4.4}$ & $28.4^{+5.3}_{-6.8}$ \\

\hline
   & JP  &$13.0^{+17.0}_{-18.0}$ &$27.0^{+15.0}_{-23.0}$ & $42^{+17.0}_{-29.0}$ \\
    GRG3 & & &  \\

   & Tribble   &$14^{+19.0}_{-13.0}$ &$29^{+15.0}_{-23.0}$ & $44^{+20.0}_{-30.0}$ \\

    \hline
    \end{tabular}
    
    \label{tab:Specages JP}
\end{table}

\section{Discussion}
\subsection{Distribution of spatially resolved spectral ages}
\label{sec: Discussion}

According to the JP model, GRG1 has a maximum age of 60 $\pm$ 11 Myr. This is the maximum pixel value, found in the lobes, in the spectral age map shown in Figure \ref{fig:GRG1 specages}. The core displays the youngest electrons (relating to GRG emission) with a value of 2.0 $\pm$ 15  Myr, i.e. consistent with a zero age. The northern hotspot shows a mean age of 25.7 $\pm$ 7.6 Myr, and the rest of the lobe is slightly older with a value of 40.0 $\pm$ 8.1 Myr, giving the average value shown in the table of $\sim 35$ Myr, signalling an injection of energy at the hotspot due to a collision with the IGM. The values here are calculated only including regions where the spectral age uncertainties are less than 30 per cent as shown in Figure \ref{fig:GRG1 specages}. This is to ensure the ages are not underestimated, as \textsc{brats} tends to assign low ages in regions where the spectral aging model is poorly constrained. The southern lobe displays similar ages to the northern lobe with an average JP age of 33 $\pm$ 7.55 Myr, consistent with the assumption that both lobes are created at the same time and are in a similar environment.   \\

Since the core of GRG2 displays a flat spectral index, it is likely dominated by synchrotron self-absorption and therefore cannot be well constrained by the power law models employed by \textsc{brats}. The spectral ages calculated in this region by \textsc{brats} are therefore unreliable, and thus have been masked out of the spectral age map in Figure \ref{fig:GRG2 specages}. The northern lobe and hotspot together have an average JP age of 22.7 $\pm$ 4.2 Myr and a Tribble age of 24.7 $\pm$ 4.4 Myr, which agree with the southern lobe ages of 25.7 $\pm$ 6.3 Myr using the JP model and 28.4 $\pm$ 6.8 Myr using the Tribble model. The maximum Tribble age is 47.0 $\pm$ 11 Myr. Both the JP and Tribble models are better constrained for GRG2 than for the other two GRGs, based on the uncertainties reported by \textsc{brats}. GRG2 also displays the most typical FRII morphology and expected spectral index distribution, implying that such properties are best modelled by \textsc{brats}.\\

GRG3, on the other hand, presents a much more interesting case. The uncertainties for GRG3 are twice as large as the other two GRGs, which is likely due to GRG3 being located towards the edge of the UHF field of view. Using the low sensitivity COSMOS$_8$ $L$-band map rather than the deeper mosaic of DR1 most likely also contributes to the uncertainties shown. Due to this, we extend our limits used for our calculations to include all pixels where the uncertainty is below 60 percent of the computed age. We do see slightly younger ages in the core (13 $\pm$ 18 Myr with the JP model) compared to the centre of the lobes (27 $\pm$ 22 Myr and 47 $\pm$ 21 Myr for the northern and southern lobes, respectively). However, the youngest ages in GRG3 are seen in the diffuse outer plasma. This is contrary to expectations, since these regions are far from the expected particle acceleration sites in the inner jets. The minimum ages in these outer plasma regions are $\sim$ 0 Myr, however, we caution that the uncertainties are even larger in this region ($\sim$ 50 Myr). We also see recently accelerated plasma surrounding UGC 05377 and conclude that these values are at least partially contaminated by the star-forming galaxy. The maximum JP age for GRG3 is 64 $\pm$ 31 Myr and the maximum Tribble age is 67 $\pm$ 36 Myr.\\

There is still, however, the matter of the aged core and only slight steepening in the lobes.  While the distribution of spectral indices and ages of GRG3 do not resemble a typical GRG there are similarities in the distribution of ages within WATs \citep{2008ApJ...682..186G,2022MNRAS.509.1837P}. Potentially, the dense environment found within both WATs and GRG3 is contributing to the distribution seen \citep{2001A&A...366...26E,2001ApJ...548..639K} and might even contribute to the large uncertainties. The slightly older core could mean that the central engine has recently switched off and the galaxy is entering a period of quiescence. However, more likely, it suggests that the injection index is instead higher than what we have assumed and so all ages have been increased. More sensitive observations at a range of frequencies are needed to constrain this degeneracy. This, however, does not affect the distribution of ages shown.

\subsection{Comparison with dynamical estimates}
\label{dynamical}

\begin{figure*}
    \centering
    \includegraphics[width = \linewidth]{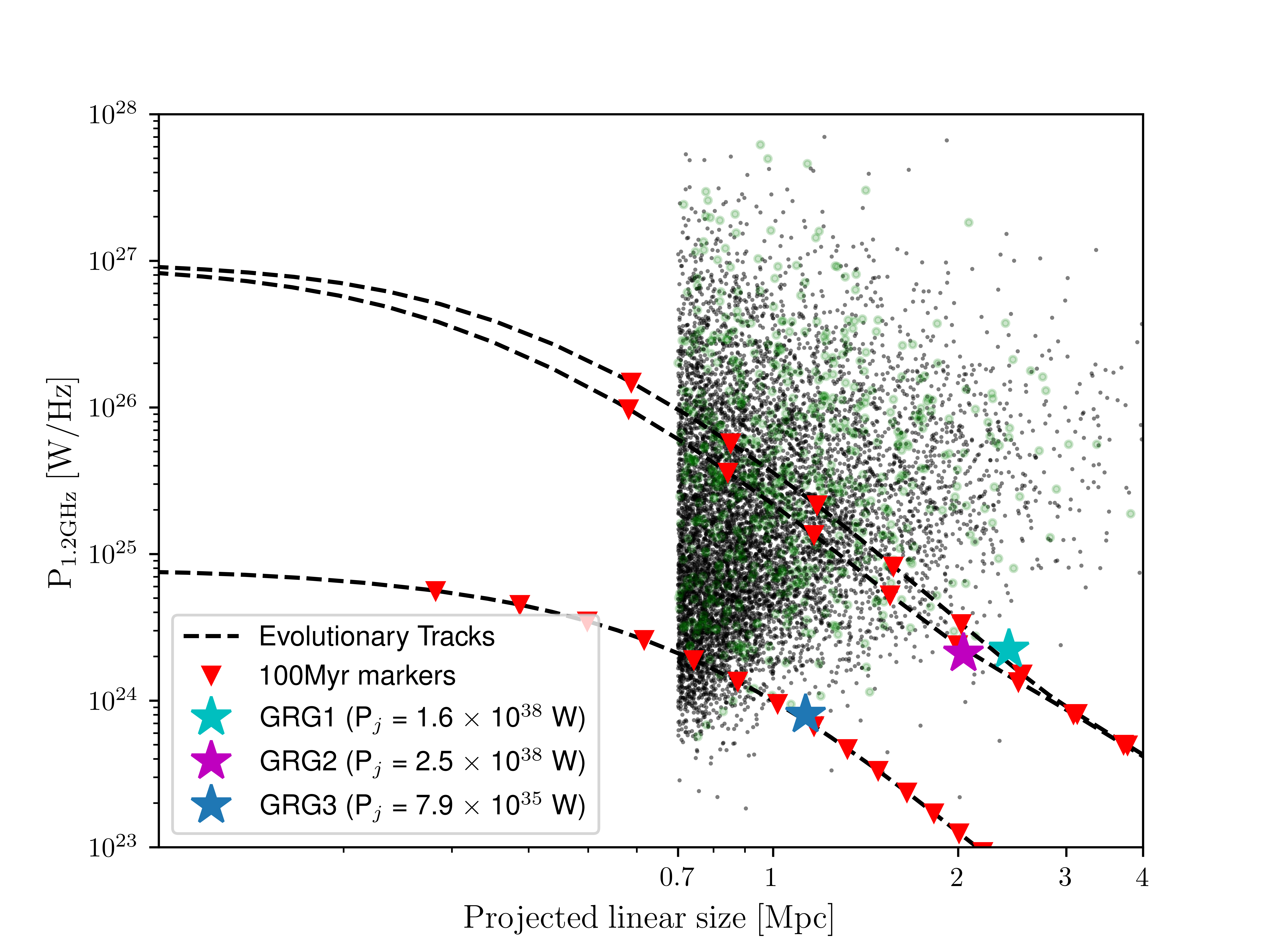}
    \caption[P-D diagram comparison]{The 1.28 GHz power-size (P-D) diagram of the COSMOS GRGs compared to literature. Objects in the compilation catalogue of \cite{Debhade...2020b} are shown as green points and sources from \cite{2023A&A...678A.151H} are shown in black. The MIGHTEE GRGs are shown as different-coloured stars and are in the lower right corner of the diagram. The black lines are evolutionary tracks generated with \cite{2018MNRAS.475.2768H}, which correspond to an idealised evolution of each GRG. The red triangles mark each hundred Myr, with the leftmost triangle on each track corresponding to 300 Myr. The jet powers for each track are given next to the corresponding GRG label.}
    \label{Fig: PD}
\end{figure*}

 We compare the power and size of the COSMOS GRGs to the those of others in the literature. Specifically, we compare to the GRG compilation catalogue of \cite{2023JApA...44...13D} and the LOTSS DR2 optically-matched source catalogue of \cite{2023A&A...678A.151H}. The latter contains accurate flux densities for many of the GRGs presented in \cite{2022MNRAS.515.2032S,oei2023measuring,2024arXiv240308037S} and \cite{2024arXiv240500232M}. The combined total is $\sim$ 9000 GRGs; see Figure \ref{Fig: PD}. The COSMOS GRGs are shown as stars and occupy the lower right corner of the diagram, showing they are still below the sensitivity limit of LOFAR, and are some of the largest and lowest luminosity GRGs detected. We use the Power-Diameter (P-D) diagram to determine the dynamical age of the galaxies as an independent comparison to the spectral age. The dynamical age is estimated based on the size, power and environment of a galaxy.  \\  

For the GRGs, we estimate the dynamical age using the \textsc{analytic}\footnote{https://github.com/mhardcastle/analytic} model from \citet{2018MNRAS.475.2768H}, shown in Figure \ref{Fig: PD} as evolutionary tracks. The \textsc{analytic} model derives a number of properties of a radio galaxy, including its power and diameter as a function of time. It assumes a density profile for the environment, as well as the power of the jets. For the GRGs, the universal pressure profile from \cite{2010A&A...517A..92A} is used, with the functional form of 

\begin{equation}
    P(r) = P_{500}\left[\frac{M_{500}}{3\times 10^{14}h_{70}^{-1}M_\odot}\right]^{\alpha_p+ \alpha'_p(x)} p(x)
\end{equation}

where $x = r/R_{500}$ and $p(x)$ is a generalised Navarro-Frenk-White profile. This profile is chosen as it has been shown to represent the pressure profile of poor cluster/big group galaxies. It also has the added advantage of only having one free parameter: $M_{500}$ - the total mass in a radius that corresponds to 500 times the critical density of the universe. \\

The density of the environment $M_{500}$ for GRG1 and GRG2 is estimated using the number density of galaxies within a 1.5 Mpc $h^{-1}$ radius. GRG1 and GRG2 reside in small groups containing 8 and 5 members, respectively, according to the zCOSMOS 20k group catalogue \citep{2012ApJ...753..121K}. Neither of these groups have enough spectroscopic redshift members ($N>$5) for accurate velocity dispersion rates, and so a lower limit of $M_{500} = 10^{13} $  M$_\odot$ is assumed \citep{2013SSRv..177..195R,2015MNRAS.453.2682I}. The $M_{500}$ value for GRG3 is taken directly from \citet{wen2012vizier} to be M$_{500}$= $9.3 \times 10^{13}$ $M_\odot$. The dynamical estimates are shown in Figure \ref{Fig: PD} as evolutionary tracks. Using these mass estimates, the best-fit tracks are determined empirically. These tracks imply that GRG1 and GRG2 are consistent with the evolution of a galaxy with a jet power of $P_{\textrm{j}}\sim$
10$^{38}$ W after $\sim$ 800 Myr and $\sim$ 700 Myr respectively. The suggested ages for GRG3 are consistent with the evolution of a  $P_{\textrm{j}} \sim  10^{36}$ W jet after $\sim$ 950 Myr.  \\

The evolution of the GRGs in \textsc{analytic} is derived assuming the jets at no point switch off. For GRG3, if we instead assume the jets have turned off recently, as potentially implied by the spectral ages in the core,  the estimated jet power would increase, and the dynamical age of GRG3 would decrease. This, however, adds an extra level of degeneracy to the model, since unless the true age before turn-off is known, there is no constraint on the jet power. Thus, GRG3 would be equally well represented by a young source with a jet power of $P_{\textrm{j}} \sim  10^{39}$ W with a dynamical age of $\sim$ 300 Myr. Hence, we simply keep the calculated dynamical age of GRG3 as an upper limit.\\

Comparing the maximum Tribble ages to their dynamical counterparts, the dynamical ages are many times larger than their spectral counterparts, though these there are some known discrepancies between spectral and dynamical age that warrant consideration.\\

The spectral ages of galaxies are mainly dependent on two parameters: the break frequency and the magnetic field. The magnetic field strength of galaxies is not trivial to determine and the distribution and evolution of the magnetic field is much more complicated than what even the Tribble model assumes. The magnetic field will decrease over time and will most likely be higher at the core and the hotspots where the electrons originate before
they permeate into the surrounding environment. This may contribute to the discrepancies between spectral and dynamical ages. \\ 

Another source of uncertainty is that the spectral age models do
not consider other loss mechanisms in the lobes, such as adiabatic
cooling, nor the effects of electron mixing. Adiabatic losses within a galaxy would
result in an overestimation of the spectral ages within the plasma
\citep{1999AJ....117..677B}. Electron mixing describes the phenomenon that even
the faintest radio emission traces electron populations of different
energies \citep{2018MNRAS.473.4179T,10.1093/mnras/stz3396}. Electron mixing
biases age towards younger estimates and is considered to be the
cause of spectral ages at sub-equipartition magnetic fields still being
at most half of their dynamical counterparts \citep{10.1093/mnras/stz3396}. \\

Furthermore, the dynamical ages are limited without true knowledge of the jet power and the proper length, though they are fairly robust against changes to these parameters. The dynamical age is however very sensitive to changes in the density of the environment, with an order of magnitude increase in density increasing the dynamical age by a factor of $\sim$ 2.   \\

Taking these limitations into account does not entirely reconcile the spectral ages with the dynamical ages, even with an estimate of the environment. Thus, we conclude similarly to \cite{2020MNRAS.491.5015M} and \cite{2023MNRAS.523..620P} that there are extra processes that either or both spectral and dynamical models do not account for. The contamination from electron mixing has not been very well constrained in GRGs and could be playing an increased role here.

\section{Summary and Conclusions}
\label{ch: conclusion}

We have presented spatially-resolved spectral index and age maps of three GRGs within the COSMOS field using MIGHTEE $L$-band data and new MeerKAT UHF band observations. MGTC J100022.85+031520.4, dubbed GRG3, is presented here for the first time. We summarise our main results here. \\

\begin{enumerate}

    \item MGTC J100022.85+031520.4, (GRG3), is a giant radio galaxy located in the northernmost part of the COSMOS primary beam in the UHF band of MeerKAT. It is hosted by an elliptical galaxy SDSS J100022.85+031520 with a redshift of $z$ = 0.1034 ± 0.00002.  It has a projected linear size of 1.29 Mpc and total power at 1284 MHz of 5.97 $\pm$ 0.02 $\times 10^{23}$ W Hz$^{-1}$. It shows similar features to a wide angle tail (WAT) galaxy.

    \item The spectral analysis was performed using two different models: the Jaffe-Perola standard model and the Tribble model. The two models agreed with each other within their uncertainties with the Tribble model consistently producing older estimates of the electron ages within the sources.
    
    \item  GRG1 was found to have the oldest electrons, with a maximum Tribble age of 68 $\pm$ 13 Myr and a mean Tribble lobe age of 34 Myr.  GRG3 has a maximum (Tribble) lobe age of $\sim$ 67 Myr, and GRG2 has the youngest maximum Tribble age of $\sim$ 47 Myr in the lobes.
    
    \item Based on the spectral age distribution, GRGs 1 and 2 are likely to still be active. The steep spectral index in the core of GRG1 indicates a restarted jet.
    
    \item The distribution of ages within the GRGs showcases the history of their evolution. GRGs 1 and 2 have been able to grow in relatively isolated environments and thus have the expected distribution of ages with the youngest in the core and the oldest in the lobes. GRG3 however, shows more complexity, possibly due to its being in a denser cluster environment and having more interactions with its surroundings.
    
    \item Dynamically, GRG3 was found to be consistent with a galaxy of $P_{\textrm{j}} \sim 10^{36}$ W jet power which is $\sim$ 950 Myr old, while GRGs 1 and 2 were consistent with 800 Myr and 700 Myr old galaxies with jet powers of $P_{\textrm{j}} \sim 10^{38}$ W. These ages are many times larger than their spectral counterparts, indicating that there are processes unaccounted for in either the spectral or dynamical age estimates.
    
    \item This work demonstrates the various limitations of the spectral age calculations and the discrepancies between the spectral ages and dynamical ages. However, it also shows that the relative ages determined through spectral ageing are valuable in determining the dynamics and evolution of GRGs through the distribution of electron ages throughout the GRG.

\end{enumerate}

\section*{Acknowledgements}

We thank the anonymous referee for useful suggestions that improved this paper. KC’s research is supported by the South African Radio Astronomy Observatory (SARAO). JD, KC, MJ and IH acknowledge support from an Africa-Oxford Catalyst Collaboration Grant (AfOx-290), which has made this research possible. JD also acknowledges support from a UCT Research Development Grant. MJJ, IHW and CLH acknowledge generous support from the Hintze Family Charitable Foundation through the Oxford Hintze Centre for Astrophysical Surveys. MJJ and IH also acknowledge support from a UKRI Frontiers Research Grant [EP/X026639/1], which was selected by the ERC, and the UK Science and Technology Facilities Council [ST/S000488/1]. FXA acknowledges the support from the National Natural Science Foundation of China (12303016) and the Natural Science Foundation of Jiangsu Province (BK20242115). I.D. acknowledges funding by the European Union - NextGenerationEU, RRF M4C2 1.1, PRIN 2022JZJBHM: "AGN-sCAN: zooming-in on the AGN-galaxy connection since the cosmic noon" - CUP C53D23001120006. The MeerKAT telescope is operated by the South African Radio Astronomy Observatory, which is a facility of the National Research Foundation, an agency of the Department of Science and Innovation. We acknowledge the use of the ilifu cloud computing facility – www.ilifu.ac.za, a partnership between the University of Cape Town, the University of the Western Cape, Stellenbosch University, Sol Plaatje University and the Cape Peninsula University of Technology. The ilifu facility is supported by contributions from the Inter-University Institute for Data Intensive Astronomy (IDIA – a partnership between the University of Cape Town, the University of Pretoria and the University of the Western Cape), the Computational Biology division at UCT and the Data Intensive Research Initiative of South Africa (DIRISA). This work made use of the CARTA (Cube Analysis and Rendering Tool for Astronomy) software (DOI 10.5281/zenodo.3377984 –  https://cartavis.github.io). Funding for SDSS-III has been provided by the Alfred P. Sloan Foundation, the Participating Institutions, the National Science Foundation, and the U.S. Department of Energy Office of Science. The SDSS-III web site is http://www.sdss3.org/.SDSS-III is managed by the Astrophysical Research Consortium for the Participating Institutions of the SDSS-III Collaboration including the University of Arizona, the Brazilian Participation Group, Brookhaven National Laboratory, Carnegie Mellon University, University of Florida, the French Participation Group, the German Participation Group, Harvard University, the Instituto de Astrofisica de Canarias, the Michigan State/Notre Dame/JINA Participation Group, Johns Hopkins University, Lawrence Berkeley National Laboratory, Max Planck Institute for Astrophysics, Max Planck Institute for Extraterrestrial Physics, New Mexico State University, New York University, Ohio State University, Pennsylvania State University, University of Portsmouth, Princeton University, the Spanish Participation Group, University of Tokyo, University of Utah, Vanderbilt University, University of Virginia, University of Washington, and Yale University. This work is based on the research supported in part by the National Research Foundation of South Africa (Grant Number 151060).The Hyper Suprime-Cam (HSC) collaboration includes the astronomical communities of Japan and Taiwan, and Princeton University. The HSC instrumentation and software were developed by the National Astronomical Observatory of Japan (NAOJ), the Kavli Institute for the Physics and Mathematics of the Universe (Kavli IPMU), the University of Tokyo, the High Energy Accelerator Research Organization (KEK), the Academia Sinica Institute for Astronomy and Astrophysics in Taiwan (ASIAA), and Princeton University. Funding was contributed by the FIRST program from Japanese Cabinet Office, the Ministry of Education, Culture, Sports, Science and Technology (MEXT), the Japan Society for the Promotion of Science (JSPS), Japan Science and Technology Agency (JST), the Toray Science Foundation, NAOJ, Kavli IPMU, KEK, ASIAA, and Princeton University.  

\section*{Data Availability}

The data underlying this article was accessed from the South African Radio Astronomy Observatory (SARAO; www.sarao.ac.za). Upon publication, the radio images from MIGHTEE DR1 will be accessible through: https://doi.org/10.48479/7msw-r692. The MeerKAT UHF-COSMOS fits image cutouts generated in this research will be shared upon reasonable request. Please contact the PI of the data (J. Delhaize; drjdelhaize@gmail.com). The raw MeerKAT visibilities for which any proprietary period has expired can be obtained from the SARAO archive at https://archive.sarao.ac.za .








\bsp	
\label{lastpage}
\end{document}